\documentclass[12pt,preprint]{aastex}
\usepackage{epsfig}
\usepackage{lscape}

\def\Msun{\hbox{\it M$_\odot$}}

\def\sgrastar{\mbox{Sgr A$^*$}}
\def\cxo{\textit{Chandra X-ray Observatory}}
\newcommand{\etal}{\mbox{et al.}}

\newcommand{\ergs}{erg s$^{-1}$}

\begin{document}
\title{Near-Infrared Counterparts to \textit{Chandra} X-ray Sources Toward the Galactic Center. {\sc I}. Statistics and a Catalog of Candidates}

\shorttitle{}

\author{J.~ C. Mauerhan\altaffilmark{1, 2}, M.~ P. Muno\altaffilmark{3}, M.~R. Morris\altaffilmark{2}, F.~ E. Bauer\altaffilmark{4},  S. Nishiyama\altaffilmark{5}, T. Nagata\altaffilmark{5}}

\altaffiltext{1}{Spitzer Science Center, California Institute of Technology, Pasadena, CA 91125, USA; mauerhan@ipac.caltech.edu}
\altaffiltext{2}{Department of Physics and Astronomy, University of California, Los Angeles, CA 90095, USA}
\altaffiltext{3}{Space Radiation Laboroatory, California Institute of Technology, Pasadena, CA 91125, USA}
\altaffiltext{4}{Columbia Astrophysics Laboratory, Columbia University, Pupin Laboratories, 550 West 120th Street, New York, NY 10027, USA}
\altaffiltext{5}{Department of Astronomy, Kyoto University, Kyoto 606-8502, Japan}

\begin{abstract}
We present a catalog of 5184 candidate infrared counterparts to X-ray sources detected toward the Galactic center. The X-ray sample contains 9017 point sources detected in this region by the {\cxo} during the past decade, including data from a recent deep survey of the central $2^{\circ}\times0\fdg8$ of the Galactic plane. A total of 6760 of these sources have hard X-ray colors, and the majority of them lie near the Galactic center, while most of the remaining 2257 soft X-ray sources lie in the foreground. We cross-correlated the X-ray source positions with the 2MASS and SIRIUS near-infrared catalogs, which collectively contain stars with a 10$\sigma$ limiting flux of $K_s\le15.6$ mag. In order to distinguish absorbed infrared sources near the Galactic center from those in the foreground, we defined \textit{red} and \textit{blue} sources as those which have $H-K_s\ge0.9$ and $<0.9$ mag, respectively. We find that $5.8\pm1.5$\% (2$\sigma$) of the hard X-ray sources have real infrared counterparts, of which $228\pm99$ are red and $166\pm27$ are blue. The red counterparts are probably comprised of Wolf--Rayet and O stars, high-mass X-ray binaries, and symbiotic binaries located near the Galactic center. Foreground X-ray binaries suffering intrinsic X-ray absorption could be included in the sample of blue infrared counterparts to hard X-ray sources. We also find that $39.4\pm1.0$\% of the soft X-ray sources have blue infrared counterparts; most of these are probably coronally active dwarfs in the foreground. There is a noteworthy collection of $\approx$20 red counterparts to hard X-ray sources near the Sagittarius B H {\sc ii} region, which are probably massive binaries that have formed within the last several Myr. For each of the infrared matches to X-ray sources in our catalog we derived the probability that the association is real, based on the source properties and the results of the cross-correlation analysis. These data are included in our catalog and will serve spectroscopic surveys to identify infrared counterparts to X-ray sources near the Galactic center. 
\end{abstract}

\section{Introduction}
Over the past decade, observations with the \textit{Chandra X-ray Observatory} have revealed a collection of $\sim$$10^{4}$ X-ray point sources within the central $2^{\circ}\times0\fdg8$ of the Galaxy, in combined observations totaling 2.25 Ms of exposure (Muno et al. 2009). This unique concentration of X-ray sources is unparalleled elsewhere in the sky by over an order of magnitude. The soft X-rays ($E<1$ keV) of more than half of them are highly absorbed, indicative of distances near the Galactic center. Stars at various evolutionary stages are represented in this population, including accreting white dwarfs in cataclysmic variables (CVs) and symbiotic binaries, massive Wolf--Rayet and O stars, and wind-accreting neutron stars and black holes in high-mass X-ray binaries (HMXBs). These exotic objects are of particular astrophysical interest, as they provide much needed information on a wide array of outstanding problems involving accretion physics, X-ray production, binary evolution, the birth of black holes, and the mode and history of massive star formation in the inner Galaxy. 

The X-ray properties alone are insufficient to sort and characterize the \textit{Chandra} population in the Galactic center; the detection of counterparts at multiple wavelengths is essential. Unfortunately, the task of identifying counterparts is complicated by the high stellar confusion in the Galactic plane and the near-complete extinction of optical and UV photons by interstellar gas and dust along the line of sight. In the infrared, multiple sources are typically found within the $\sim$$1\arcsec$ error circle of \textit{Chandra} (e.~g., see Bandyopadhyay et al. 2005), and suffer spatially variable extinction that renders the determination of intrinsic stellar colors on the Rayleigh--Jeans tail practically impossible. Therefore, the spectral types of counterparts cannot be determined photometrically. However, progress in determining the nature of infrared counterparts to the Galactic center X-ray population can be made by approaching the sample statistically. Bright infrared sources are less likely to randomly fall within the \textit{Chandra} error circles than more numerous faint stars. Thus, if the \textit{Chandra} population contains luminous objects such as massive stars and HMXBs, then there should be an excess of bright infrared counterparts within the error circles of X-ray sources, in comparison to a randomly sampled field in the Galactic center. 

Laycock et al. (2005) searched for near-infrared counterparts to X-ray sources in the inner {10\arcmin} around \sgrastar, and determined that the number of hard \textit{Chandra} sources from the catalog of Muno et al. (2003) with infrared matches having $K_s<15.5$ mag is statistically equivalent to the number of random interloping field stars. Arendt et al. (2008) also found no significant correlation between the same X-ray sources and mid-infrared stars from \textit{Spitzer}/IRAC surveys.  Nonetheless, Muno et al. (2006), Mikles et al. (2006), Mauerhan et al. (2007), and Hyodo et al. (2008) have collectively identified five new X-ray-emitting Wolf-Rayet and O stars by carefully selecting infrared matches to X-ray sources for spectroscopic follow-up, demonstrating that although accidental coincidences dominate the sample, there are genuine detectable counterparts to be found. 

Taking advantage of the substantial astrometric improvement that recent \textit{Chandra} surveys of the central Galaxy have provided, we present a statistically significant detection of several hundred infrared counterparts to hard \textit{Chandra} X-ray sources. The resulting catalog of candidate counterparts will be of use in spectroscopic surveys that will identify and characterize members of the X-ray population near the Galactic center. In Section 2, we describe our X-ray and infrared samples. In Section 3, we describe the details of our cross-correlation and probability analysis, and present the resulting statistics and the catalog of all candidate infrared counterparts to the X-ray sample. In Section 4 we compare our results to other surveys, describe the various stellar exotica contained within our sample, and discuss prospects for future experiments.

\section{The Sample}
\subsection{X-ray Sources and Hardness Selection }
For the X-ray sample, we used the \textit{Chandra} catalog of 9017 X-ray point sources detected in a $2^{\circ}\times0\fdg8$ field around the Galactic center in $l\times b$, respectively (Muno et al. 2009). This catalog incorporates all \textit{Chandra} ACIS-I observations of this region performed through the end of 2007 July, totaling 2.25 Ms of exposure for all combined fields. It includes 1 Ms of observations of the central 20 pc around Sgr A*, 30 shallow (12 ks) exposures from the Wang et al. (2002) survey, several deep pointings toward the Arches and Quintuplet region and the Sgr B and Sgr C H {\sc ii} regions, and 15 relatively recent 40 ks exposures that were obtained to improve the sensitivity and astrometric accuracy of the catalog from Wang et al. (2002). At the distance to the Galactic center of 8 kpc, the survey field covers a projected physical area of $\approx280\times 110$ pc, and is sensitive to X-ray sources with luminosities of $4\times10^{32}$ {\ergs} (0.5--8.0 keV) over 1 deg$^2$ of area (90\% confidence), and is an order of magnitude more sensitive in the deepest exposure toward the {\sgrastar} field. The inclusion of recent deep observations has significantly improved the photon statistics and thus provides a substantial astrometric improvement over previous Galactic center catalogs (e.g., Muno et al. 2003); 60\% of the sources have relative astrometry accurate to $<$1\arcsec, and 20\% are accurate to $<$0\farcs5.  The absolute astrometry is accurate to 0\farcs1--0\farcs2 for all the deep 20-ks fields, and $\approx$0\farcs6 for the shallower fields.  

Foreground X-ray sources are separated from more distant ones by their colors, which are quantified by calculation of their hardness ratio, given by HR=$(h-s)/(h+s)$, where $h$ and $s$ are the fluxes in the hard and soft energy bands, respectively. For the soft color, HR0, $h$ and $s$ are the fluxes in the 2.0--3.3 keV and 0.5--2.0 keV bands. Soft X-ray photons are sensitive to absorption by interstellar gas and dust; so, in general, only foreground objects should be significant sources of soft X-ray photons. We follow the criteria in Muno et al. (2009), where \textit{soft sources} are defined as those having $-1.0\le\textrm{HR0}<-0.175$, which corresponds to a hydrogen absorption column of $N_{\textrm{\scriptsize{H}}}<4\times10^{22}$~cm$^{-2}$. There are 2257 such soft sources. X-ray sources that are located near or beyond the Galactic center were selected as those that either have $\textrm{HR0}\ge$$-0.175$, consistent with $N_{\textrm{\scriptsize{H}}}>4\times10^{22}$~cm$^{-2}$, or that were completely extinguished in the soft 0.5--3.3 keV energy bands. These are defined as \textit{hard sources}, and there are 6760 of them. Of these, 3600 are completely extinguished at photon energies below 3.3 keV. 

For X-ray sources with $\textrm{HR0}=-1$, the classification of these as soft sources is reliable. The same goes for hard sources that are only detected at photon energies above 3.3 keV. However, for sources with intermediate values of HR0, classifying them as hard or soft is more difficult. The uncertainties in HR0 are significant; although a given source may have an HR0 value consistent with the hard or soft criterion, the 90\% confidence envelopes of HR0 frequently have upper or lower bounds that extend into the opposite regime. For example, $\approx$47\% of  the soft X-ray sources have HR0 upper limits that extend into the hard regime. Similarly, $\approx$10\% of the hard sources have HR0 lower bounds that extend into the soft regime. Thus, we may expect that sources lying near the boundary of our HR0 cutoff are classified as hard or soft less reliably. This uncertainty might be important for sources lying at intermediate distances between Earth and the Galactic center. 

\subsection{Infrared Sources and Color Selection} 
The main near-infrared catalog used for this work was produced by the Simultaneous-3-color InfraRed Imager for Unbiased Surveys (SIRIUS; Nagashima et al. 1999, Nagayama et al. 2003) on the 1.4 m Infrared Survey Facility (IRSF) in South Africa. The SIRIUS survey covered the region $|l|\lesssim2^{\circ}$ and $|b|\lesssim1^{\circ}$ with a pixel scale of 0\farcs45. The typical seeing during observations was 1\farcs2 full-width-half-maximum in the \textit{J} band, and the average 10$\sigma$ limiting Vega magnitudes in the survey were $J$=17.1, $H$=16.6, and $K_s$=15.6 (see Nishiyama et al. 2006). The SIRIUS catalog contains $\approx2\times10^6$ stars within the $2^{\circ}$$\times$$0\fdg8$ field covered by our X-ray catalog and is 2--3 mag more sensitive than the Two Micron All Sky Survey (2MASS; Cutri et al. 2002), mainly because the finer pixels of SIRIUS mitigate some of the source confusion in the Galactic center region. However, the brightest SIRIUS sources ($K_{s}\lesssim8$ mag) are saturated, so we also correlated our X-ray catalog with 2MASS, using sources with $K_{s}<8.0$ mag. In general, the uncertainty on the near-infrared positions were $\lesssim0\farcs1$ and $\lesssim0\farcs3$, for the SIRIUS and 2MASS sources, respectively. Thus, for the cross-correlation to follow in Section 3, we set $\sigma_{\textrm{\scriptsize{IR}}}$ to these values. Based on the quoted positional uncertainties for infrared and X-ray sources, it is clear that the relative astrometric errors of the X-ray sources dominate the positional uncertainties of X-ray/infrared matches.

In order to distinguish foreground infrared sources from absorbed sources near the Galactic center, we established a simple selection criterion based on infrared color. 
In Section 2.1, hard X-ray sources located near or beyond the Galactic center were selected as those having HR0 values consistent with $N_{\textrm{\scriptsize{H}}}>4\times10^{22}$~cm$^{-2}$. Using the relation of Predehl \& Schmitt (1995), which gives $N_{\textrm{\scriptsize{H}}}/A_{V}=1.8\times10^{21}$ cm$^{-2}$ mag$^{-1}$, the adopted $N_{\textrm{\scriptsize{H}}}$ value corresponds to an optical extinction of $A_{V}\lesssim22.3$ mag. We derived the corresponding extinction in the \textit{J}, \textit{H} and $K_s$ bands using the ratios reported in Nishiyama et al. (2008) for Galactic bulge fields ($A_V$:$A_J$:$A_H$:$A_{K_s}$=1.0:0.188:0.108:0.062). We determined that with a visual extinction of $A_{V}>22.3$ mag, the bluest O stars from Martins \& Plez (2006) should have $J-H\ge1.7$ and $H-K_s\ge0.9$ mag.  We define \textit{red} sources as those having $H-K_s$ values above this color restriction, which amounts to $\approx1.8\times10^6$ sources. We ignore the $J-H$ colors, since $\approx$80\% of the infrared sample have no $J$-band detections.  However,  $\approx$10\% of the infrared sources have only $K_s$-band detections. Based on our $H$-band sensitivity, this automatically implies $H-K_s>1$ mag, which is redder than the cutoff we set for the bluest possible stars. Thus, we also included $K_s$-only detections in the sample of red sources.  Alternatively, we defined \textit{blue} sources as those with $H-K_s< 0.9$ mag, which amounts to $\approx1.7\times10^5$ sources. Thus, approximately 90\% of the infrared sources with $K_s\le15.6$ mag satisfy the red criterion and 10\% satisfy the blue criterion. Finally, $\approx$0.2\% of the infrared sample (4279 sources) were detected in the $J$ and $K_s$ bands, but not in the $H$ band. Although these sources comprise a statistically insignificant fraction of the infrared sample, we included them in the interest of completeness. The color criterion for the null-$H$ sources was obtained by combining the $J-H$ and $H-K_s$ color restrictions calculated above, which corresponds to $J-K_s\ge2.6$ and $<2.6$ for red and blue sources, respectively. 
 
\section{Analysis and Results}
\subsection{Cross-correlation of X-ray and Infrared Sources}
Candidate infrared counterparts to X-ray sources, which we call \textit{matches}, were selected by identifying infrared sources having angular separations from the nominal X-ray positions that are smaller than the quadrature sum of the positional uncertainties of the infrared and X-ray sources. To determine the number of real infrared matches, which we call \textit{counterparts}, we must account for the frequency of accidental matches. We did this by first counting the number of matches that occurred between astrometrically aligned catalogs, $N_{\textrm{\scriptsize{align}}}$. The catalogs were then shifted by a random amount in R.A. and decl., and the number of matches that occurred randomly, $N_{\textrm{\scriptsize{ran}}}$, was counted.  The random positional shifts were set to be uniformly distributed within an angular radius of 10 times the largest X-ray error circle in the catalog, which prevented shifts from extending beyond $\approx$1\farcm3.  The shifting and counting process was iterated 100 times, yielding an average, $N_{\textrm{\scriptsize{ran}}}$, and standard deviation, $\sigma$, for the number of random matches. This process was performed incrementally with respect $K_s$ magnitude, so that the significance of matches could be assessed as a function of stellar brightness. 

We performed an initial cross-correlation of all X-ray and infrared sources, imposing no X-ray hardness or infrared color constraints. The soft X-ray and infrared color distribution of the resulting matches is illustrated in Figure 1. We performed this exercise in order to assess the suitability of our X-ray and infrared selection criterion described in Section 2, which are justified by Figure 1. With the X-ray hardness and infrared color constraints imposed, separate cross-correlations were performed for four potential combinations of hard and soft X-ray sources, and blue and red infrared sources. In Table 1 we list the results of the cross-correlation of hard X-ray sources with red and blue infrared sources. The table includes the number of observed matches listed as a function of limiting $K_s$ magnitude, along with the number of random matches and the standard deviation from the shift-and-match simulations. The final two columns in the table list the expected number of real infrared counterparts, $N_{\textrm{\scriptsize{real}}}$, and the significance of  the detection of true counterparts, expressed as $N_{\textrm{\scriptsize{real}}}/\sigma$. These results are illustrated graphically in Figure 2. 

Out of 2575 red infrared matches to hard X-ray sources, we find that there are $228\pm99$ (2$\sigma$, used hereafter) real counterparts with $K_s\le15.6$ mag, which amounts to $3.4\pm1.5$\% of the 6760 hard sources. The largest increases in significance occur at $K_s=7.0$--8.5, 11--11.5, and 12--12.5 mag.  Most of the red matches to hard X-ray sources probably lie near the Galactic center, although there may also be foreground sources in this sample that are embedded in foreground regions of locally high extinction. Alternatively, we find $166\pm28$ real, blue infrared counterparts to hard X-ray sources, out of 340 matches, which amounts to $2.5\pm0.4$\% of the hard sources. The lower occurrence of random matches between this group of infrared/X-ray sources results in their relatively high statistical significance. Many of these sources probably lie at intermediate distances, or are foreground sources that suffer intrinsic soft-X-ray absorption. In total, there are $394\pm92$ real infrared counterparts to hard X-ray sources, which amounts to $5.8\pm1.5$\% of them. 

In Table 2 we list the results of a similar analysis for the soft X-ray sources, and illustrate the results in Figure 3. There are $890\pm23$ (2$\sigma$) blue infrared counterparts within the sample of 1007 blue matches. This sample is dominated by foreground objects that suffer relatively little absorption at infrared and X-ray wavelengths. Finally, we performed a cross-correlation between soft X-ray sources and red infrared sources. We did not expect to obtain a significant number of matches in this case, since red infrared sources, in general, should be associated with hard X-ray sources; the results in Table 2 confirm this. However, an anti-correlation between red-infrared and soft-X-ray sources emerges for $K_s\ge14.5$ mag, which becomes more significant with increasing values of $K_s$. We offer the following explanation. Dark obscuring clouds lying at intermediate distances between Earth and the Galactic center will obscure more distant sources nearer to the Galactic center, while the soft X-ray sources in the foreground can only lie in front of such clouds. Thus, in our simulations, the chance of a random match between soft X-ray sources and red infrared sources is higher when the sources are randomly shifted in position; soft X-ray sources in front of obscuring clouds are shifted into lines of sight that penetrate deeper into the Galaxy and, hence, contain more faint red sources. To confirm this, we performed the cross-correlation using only soft X-ray sources with the most accurate positions ($\sigma_{\textrm{\tiny{X}}}\le1{\arcsec}$). Random positional shifts of $>5{\arcsec}$ significantly increased the chance of achieving a random red match to a soft X-ray source. Shifts smaller than this produced a number of random matches that was similar to $N_{\textrm{\scriptsize{align}}}$. We note that only this particular combination of red infrared and soft X-ray sources were found to have a positional bias in the number of random matches.

We also performed cross-correlation experiments by assigning uniform 2\arcsec\ or 3\arcsec\ positional uncertainties to the X-ray sources. A similar number of real counterparts was obtained, although many more spurious matches were also found.  Alternatively, using smaller matching radii results in fewer spurious matches, but also fewer genuine matches. This test provides some validation of the uncertainties on the positions of the X-ray sources claimed by Muno et al. (2009), although the large number of random matches when using large matching radii prevents us from empirically determining the positional uncertainties.

\subsection{Probabilities for Individual Matches}
In the interest of spectroscopic observations of potential counterparts, it is important to assess the probability that a given match is real, so that the large number of targets can be prioritized quantitatively. We adopt a method similar to that used by Bauer et al. (2000) to investigate radio and optical matches to extragalactic X-ray sources. In this formalism, the probability that a given match is real is expressed as a function of (1) the angular separations between the X-ray sources and their infrared counterparts, which we call $\phi$, (2) the fraction of X-ray sources that actually have detectable infrared counterparts, which we call $f$, and (3) the local stellar surface density in the background, which we call $\rho$. 

The distribution of angular separations between X-ray sources and their true counterparts can be expressed as a Gaussian, given as a normalized probability by
\begin{equation}
p(\phi) = \frac{1}{2\pi\sigma^2} \exp\left( - \frac{\phi^2}{2\sigma^2} \right)
\end{equation}
\noindent{where} $\sigma^{2}=\sigma_{\textrm{\scriptsize{X}}}^{2}+\sigma_{\textrm{\scriptsize{IR}}}^{2}$, and $\sigma_{\textrm{\scriptsize{X}}}$ and $\sigma_{\textrm{\scriptsize{IR}}}$ are the uncertainties in the X-ray and infrared positions, respectively. The fraction of X-ray sources with detectable infrared counterparts, $f$, is obtained by dividing the $N_{\textrm{\scriptsize{real}}}$ columns in Tables 1 and 2 by the total number of hard (6760) or soft (2257) X-ray sources. For a given infrared match to an X-ray source, the value of  $\rho$ was computed for red or blue sources (see Section 2.2) by counting all of the respective stars within an angular radius defined by the average value of the 100 positional offsets performed in the shift-and-match routine described in Section 3.1, which was typically $\approx$30{\arcsec}.  Since in our case the probability is also dependent on infrared brightness (bright infrared matches to X-ray sources are found less frequently than faint matches), we have imposed a limiting $K_s$-magnitude dependence on $f$ and $\rho$. For the following mathematical expressions,  $K_s$ magnitude values will be expressed as $K$, in order to improve their readability. With a stellar magnitude dependence imposed, $f$ and $\rho$ become $f(K)$ and $\rho(K)$. Thus, the resulting expression for the probability that an $i^{\textrm{\scriptsize{th}}}$ match to an X-ray source is real, among $m$ total matches of the same infrared color class (i.~e., red or blue), is proportional to the product of $f(K_i)$, $p(\phi_i)$, and $1/\rho(K_i)$:
\begin{mathletters}\label{peq}
\begin{eqnarray}
P_i = c \times f({K_i}) \times \frac{p(\phi_i)    } {{\rho}({K_i}) }~\mbox{~~~~if}~~i>0
\end{eqnarray}
\begin{eqnarray}
P_0=c~(1-f_{\textrm{\scriptsize{lim}}})
\end{eqnarray}
\end{mathletters}
\noindent{where} $f({K_i})$ is the fraction of X-ray sources that have matches with $K\lesssim K_{i}$, and ${ \rho}(K_i)$ is the background density of infrared sources that have $K\lesssim {K_{i}}$. These two values are approximate because for Tables 1 and 2 we calculated ${f}(K_i)$ at intervals of 0.5 mag, ranging from $K=6.5$ to 15.0 mag, and at our limiting magnitude of 15.6. Since the functional behavior of $f({K})$ is smooth, for each infrared match we used the $f({K})$ value at the nearest $K$ magnitude point. For example, for infrared matches having $K=11.1$ and 11.4 mag, we used $f({K})$ for $K\le11.0$ and 11.5 mag, respectively. $P_0$ is the probability that none of the sources is the true counterpart, and $f_{\textrm{\scriptsize{lim}}}$ is the fraction of X-ray sources having infrared counterparts at our limiting magnitude of $K\le15.6$. The constant $c$ is generated by the assumption that, for a given X-ray source, the sum of probabilities over all $m$ candidates, including $P_0$, is unity: 
\begin{equation}
c^{-1}=\sum_{i=1}^{m}\frac{f(K_i)} { \rho(K_i)} ~p(\phi_i)+(1-f_\textrm{\scriptsize{lim}}),
\end{equation}

All candidate red and blue infrared counterparts to hard and soft X-ray sources are listed in Table 3, which includes infrared photometry, positions, and all of the parameters necessary to calculate the individual match probabilities, when combined with $f$ and $f_{\textrm{\scriptsize{lim}}}$, which can be obtained from Tables 1 and 2. Cases involving multiple infrared matches to a given X-ray source are sorted by probability.  
Figure 4 illustrates the probability distribution of three sets of X-ray and infrared matches: red infrared with hard X-ray sources, blue infrared with hard X-ray sources, and blue infrared with soft X-ray sources. Owing to the anti-correlation of faint, red matches to soft X-ray sources, which results in negative values of $f_\textrm{\scriptsize{lim}}$, as described above, we did not calculate probabilities for these particular matches. 

The probabilities for the 1007 blue infrared matches to soft X-ray sources have an average and standard deviation of 59\% and 27\%, respectively. These relatively high probabilities are the result of both the lower occurrence of random matches between blue infrared and soft X-ray sources,  and the fact that true counterparts within this sample lie in the foreground, in which case their infrared emission, and their less-absorbed soft X-ray emission, are more readily detectable than that of the more distant sources. Indeed, for most of the 2575 red matches to hard X-ray sources, the probability that a given match is genuine is quite low, having an average probability and standard deviation of 7\% and 10\%, respectively. Exceptions include relatively bright infrared matches to X-ray sources having small error circles.  For example, infrared matches to X-ray sources with $\sigma_{\textrm{\scriptsize{X}}}\le0\farcs5$ have an average probability of $\approx$20\%, and the brightest of these, having $K_s\le12.0$ mag, have an average probability of $\approx$50\%. Some previously identified sources in the Galactic center are in this group, including Wolf--Rayet and O-supergiant X-ray sources within the Arches and Quintuplet clusters (e.g., see Wang et al. 2006) and in other Galactic center fields (Muno et al. 2006; Mikles et al. 2006; Hyodo et al. 2008). We list them in Table 4. These sources also produce the bump near $K_s=8$ mag in the upper right panel of Figure 2, and explain the increase in significance of matches having $K_s=7.0$--8.5 mag in Table 1. Owing to the rarity of Wolf--Rayet and O spectral types relative to the average late-type field star, the associated X-ray sources are almost certainly their true counterparts. However, according to our probability calculations, which are blind to spectral type, these stars have associated probabilities in the range $P\approx$20\%--80\%. These values provide some indication of the probability range one might expect for similar, currently unidentifed massive stars in the Galactic center region.  Finally, blue matches to hard X-ray sources have relatively high probabilities, compared with the red matches. This is partially due to the lower value of $\rho$ for blue stars relative to red stars, but might also result from errors in the HR0 values we used to define the hard and soft criteria in Section 2.1. These errors will be relevant for X-ray sources near the cutoff. For example, 88\% of soft X-ray sources with blue matches satisfy the soft criterion with $>90$\% confidence; 91\% of hard X-ray sources with red matches satisfy the hard criterion with $>90$\% confidence, while for those with blue matches, only 57\% satisfy the hard criterion with $>90$\% confidence.
  
\subsection{Spatial Distribution}
The apparent spatial distribution of X-ray sources and their infrared matches is illustrated in Figures 5 and 6 for hard and soft X-ray sources, respectively. In the case of red infrared matches to hard X-ray sources, our ability to determine their true spatial distribution is limited by both the high occurrence of random matches and the X-ray point-source incompleteness across the survey area. The \textit{Chandra} integration times varied for different pointings within the $2^{\circ}\times0\fdg8$ survey field, and the \textit{Chandra}/ACIS-I detector is less sensitive to sources that fall closer to the edge of its field of view.  Thus, although the X-ray sources in Figures 5 and 6 may appear clustered, this is not necessarily representative of their true distribution. However, there are several locations where the clustering of X-ray sources is undoubtedly real, at $(l, b)\approx(0\fdg0, 0\fdg2)$, $(0\fdg3, -0\fdg2)$, and $(-0\fdg35, -0\fdg25)$; in the former two regions the clustering of blue infrared matches to both hard X-ray sources (Figure 5, \textit{lower panel}) and soft X-ray sources (Figure 6, \textit{middle panel}) is obvious, while in the latter region the clustering appears to be real, albeit less extreme. In all cases the infrared matches to X-ray sources in these regions are predominantly blue, which indicates that these clusters of sources lie in the foreground.

In the remainder of our survey area, the true spatial distribution of X-ray sources and their infrared counterparts is much more difficult to ascertain. According to Muno et al. (2009), only sources with a photon flux of $F_{X}\gtrsim2\times10^{-6}$ cm$^{-2}$ s$^{-1}$ in regions where the 50\% completeness limit was below this flux threshold were securely detected across the entire survey area. This amounts to only $\approx$480, and $\approx$210 X-ray sources that satisfy the hard and soft criteria, respectively.  If we apply the same X-ray flux restriction to our infrared/X-ray matches, we are left with only 61 and 93 hard and soft matches, respectively, which is an insufficient number of sources to determine their true distribution. However, Muno et al. (2009) were able to detect a $\approx$2.8$\sigma$ excess of hard X-ray sources in the inner few arcminutes around {\sgrastar}, and near the area occupied by the Arches and Quintuplet stellar clusters. Thus, we can at least examine the statistics of red infrared and hard X-ray matches in these particular regions, and compare the results to those from other fields to determine if there is a significantly higher number of infrared counterparts in one region versus another.

We performed cross-correlations of hard X-ray and red infrared sources in four sub-fields of interest: the central `cusp' of X-ray sources centered on {\sgrastar},  a field containing the Arches and Quintuplet stellar clusters, and the Sgr B and Sgr C H {\sc ii} regions. The selected areas are illustrated by the large circles contained in the top panel of Figure 5, which all have angular radii of 8\arcmin. The X-ray sample surrounding the {\sgrastar} field was defined using the position of {\sgrastar} itself ($\alpha=266.41726$, $\delta=-29.00798$ degrees, J2000). A total of 3336 X-ray sources lie within this area, comprising over one-third of the entire catalog.  The results of the cross-correlation are listed in Table 5.  There are $3.7\pm1.3$ (2$\sigma$) real counterparts with $K_s\le8.5$ mag, and $12.0\pm8.0$ infrared counterparts with $K_s\le10.5$ mag. This number is consistent with the amount of previously identified Wolf-Rayet and O stars in this region that have already been associated with X-ray sources (e.~g., see Muno et al. 2006 and Mikles et al. 2006). We also noted the infrared/X-ray counterpart associated with source GCIRS 13 of the central cluster. Although this source has been associated with a very compact cluster of emission-line stars, revealed by high-resolution imaging observations (Maillard et al. 2004), it was detected as a single point source in our lower-resolution infrared catalog.  There was no significant detection ($N_{\textrm{\scriptsize{real}}}/{\sigma}\ge3$) of fainter counterparts in the region surrounding {\sgrastar}. However, we also performed an additional cross-correlation experiment, only using X-ray sources lying within 2{\arcmin} of {\sgrastar} that have $\sigma_{\textrm{\scriptsize{X}}}\le0\farcs6$. We detect $23\pm15$ counterparts with $K_s=14.0$--14.5 mag. This was the only magnitude range for which $N_{\textrm{\scriptsize{real}}}/{\sigma}\ge3$. Although this detection is near the limit of significance, it suggests there are at least a few real counterparts in this brightness range. 

For the cross-correlation of sources near the Arches and Quintuplet region, we used the 656 X-ray sources lying within 8{\arcmin} of a point approximately between the clusters ($\alpha=266.50761$, $\delta=-28.82549$). The results are also listed in Table 5. We found $2.9\pm0.7$ and $5.8\pm3.8$ counterparts with $K_s\le8.0$ and $K_s\le10.0$ mag, respectively. These numbers are consistent with the number known X-ray-emitting Wolf--Rayet and O stars within and near the clusters (Wang et al. 2006; Mauerhan et al. 2007). Thus, we expect that, at most, only a few similar objects remain unidentified in this region. There was no population of fainter counterparts with $N_{\textrm{\scriptsize{real}}}/{\sigma}\ge3$ in this region. 

The most significant collection of unidentified counterparts appears to be associated with  the Sgr B H {\sc ii} region. For the cross-correlation in this region we used all 253 X-ray sources lying within 8{\arcmin} of $\alpha=266.78070$ and $\delta=-28.44160$. This position corresponds to the \textit{Chandra}/ACIS-I pointing center of the deepest observation of this field (Muno et al. 2009, their Table 1). The results are listed in Table 6. We found $17\pm5$ ($N_{\textrm{\scriptsize{real}}}/{\sigma}=7.1$) counterparts with $K_s\le12.5$ mag. This high level of significance for red matches to hard X-ray sources is unparalleled elsewhere in the survey field. Thus, the apparent clustering of red matches to hard X-ray sources at $(l,b)\approx(0\fdg6, -0\fdg1)$ may, in part, be the result of a true concentration of luminous X-ray-emitting stars near Sgr B (see Figure 5, \textit{middle panel}) .

Finally, we performed a cross-correlation using the 490 X-ray sources that lie within 8{\arcmin} of the Sgr C H {\sc ii} region ($\alpha=266.08819$, $\delta=-29.43665$). With the exception of one very bright counterpart having  $K_s=6.2$ mag, we found no significant detection of counterparts fainter than this. The Sgr C region might not be as endowed with luminous X-ray emitting stars as the Arches and Quintuplet region and Sgr B. Alternatively, Sgr C might contain such stars that lie behind thick columns of obscuring material, preventing the detection of their near-infrared signal. The latter possibility would not be surprising, given that several dark obscuring clouds appear to intersect the Sgr C region. By comparison, the Sgr B field also contains dark obscuring clouds, but the majority of infrared matches we find there lie adjacent to them, in regions of lower extinction, that are favorably placed for near-infrared detection. The superposition of X-ray-emitting stars and dark obscuring clouds may not be so favorable near Sgr C.

\section{Discussion}
The modest fraction of hard \textit{Chandra} sources with infrared counterparts is in agreement with previous studies, which found that less than 10\% of the hard X-ray sources within $\approx$5\arcmin of {\sgrastar} have counterparts with $K_s<15$ mag (Laycock et al. 2005). This is consistent with the interpretation that the hard X-ray population is dominated by CVs (Muno et al. 2003), whose K--M donor stars should have $K_s\gtrsim20$ mag if they lie near the Galactic center, 4--5 mag fainter than our $K_s$-band sensitivity limit.  This conclusion is valid for both the stellar concentration around {\sgrastar}, and the extended survey area. Other objects contributing to the infrared-faint population include low-mass X-ray binaries,  and possibly some isolated neutron stars and black holes accreting material from within relatively dense regions of the interstellar medium (Agol \& Kamionkowski 2002). Relative to CVs, these objects should be significantly less numerous, since they are the products of massive-star evolution, and involve rare astrophysical circumstances. In any case, our results confirm that over 95\% of the hard X-ray sources near the Galactic center lack infrared counterparts above the sensitivity limit of our survey. 

Nonetheless, our work confirms the existence of $228\pm99$ hard X-ray sources with red infrared counterparts, many of which probably lie near the Galactic center. Several bright counterparts have already been identified as massive Wolf--Rayet and O stars (Yusef-Zadeh et al. 2002; Wang et al. 2006; Muno et al. 2006; Mikles et al. 2006; Mauerhan et al. 2007). The emission of hard X-rays from these stars, which is not a pervasive feature of single massive stars, is interpreted as a signature of binarity, where hard X-rays are generated by colliding supersonic winds or by accretion onto a compact companion. Several of these sources were found near, but not within, the Arches and Quintuplet clusters and may have originated within them (Mauerhan et al. 2007). Several others do not appear to be associated with a known stellar cluster (see Muno et al. 2006 and Mikles et al. 2006), and may be the products of an alternate, relatively isolated mode of massive star formation occurring in the Galactic center region, in tandem with the formation of dense stellar clusters. 

We expect that massive stars are likely to be responsible for the significant number of bright red infrared matches to the hard X-ray sources in the Sgr B region. Indeed, this area contains the most massive molecular cloud known in the Galaxy, in which over 60 ultra-compact H {\sc ii} regions are embedded (Gaume et al. 1995; de Pree et al. 1998). Naturally, this is a site of vigorous, massive star formation. Infrared spectroscopy of red matches in this region will likely reveal hot supergiants, similar to those already identified near the Arches and Quintuplet clusters.  The identification of such stars near Sgr B will constrain the history and mode of massive star formation in this region.

The nature of the relatively  faint counterparts in our survey, having $K_s\approx$12--15.6 mag, is currently a mystery. Population synthesis models suggest that there could be $\sim1000$ neutron stars and black holes accreting from the winds of companions with $M>3\Msun$ in our field (Pfahl et al. 2002). Assuming an average extinction of $A_{K_s}=2.5$ mag for sources near the Galactic center, the limiting magnitude of $K_s=15.6$ should enable the detection of main-sequence stars with spectral types earlier than B1{\sc V}--B3{\sc V}. Most known HMXBs have donor stars within a narrow range of spectral types from 09{\sc V}--B2{\sc V}, and with a peak at B0{\sc V} (Negueruela 1998), so we should be sensitive to similar HMXBs lying near the Galactic center. If there really are $\sim$100 or more HMXBs in our sample of infrared matches to X-ray sources, then spectroscopic confirmation of them will double the number of known Galactic HMXBs (Liu et al. 2006). A large sample of HMXBs all lying at a similar distance will elucidate the formation and evolution of these objects. For instance, given the number-flux distribution of the Galactic center X-ray sample from Muno et al. (2009), confirmed HMXBs in this sample would have low luminosities relative to most known HMXBs, and hence, low accretion rates. It has been suggested that strong magnetic fields in neutron-star HMXBs may prevent captured material from reaching the neutron star's surface, which could prevent weakly accreting HMXBs from being detected as X-ray sources (e.g., Liu \& Li 2006). The confirmation of low-luminosity HMXBs in our sample would provide the means to study this process.

In addition, our catalog of infrared matches should also be sensitive to symbiotic binaries containing late-type giant donors to accreting white dwarfs.  An M0III giant ($M_K=-4.3$ mag; Tokunaga 2000) in the Galactic center would have $K\approx12.8$ mag, well within our sensitivity limit. We should also expect symbiotics to exist at intermediate distances in the Galactic-center foreground. X-rays from accreting sources may suffer intrinsic photoelectric absorption if the accretion flow obscures the X-ray source, a property not uncommon for symbiotic binaries (e.~g., see Masetti et al. 2007). Such objects might contribute to the sample of blue infrared counterparts to hard X-ray sources. With regard to spectroscopic follow-up observations, symbiotic binaries will only be distinguishable from interloping late-type field stars by the successful detection of an infrared accretion signature, such as a Br$\gamma$ emission line of H {\sc i} at $\lambda2.17$ {$\micron$}. 

With regard to foreground X-ray sources, the tentative discovery of three clusters of blue matches to hard and soft X-ray sources at $(l, b)\approx(0\fdg0, 0\fdg2)$, $(0\fdg3, -0\fdg2)$, and $(-0\fdg35, -0\fdg25)$ is the most interesting result. All three of these regions appear to be associated with dark clouds that are relatively devoid of stars in the $K_s$-band images in Figures 5 and 6. In addition, \textit{Spitzer} $\lambda$8 {\micron} images show that these regions are associated with concentrations of bright emission from warm polycyclic aromatic hydrocarbons (PAH), which highlight the edges of the dark obscuring clouds (e.~g., see Stolovy et al. 2006, and compare their Figure 1 with our Figures 5 and 6).  Thus, these are likely to be foreground regions of active star formation, and the associated clusters of X-ray sources are probably young dwarf stars with active coronae (e.~g., see Feigelson et al. 2004). The tentative discovery of these three clusters illustrates their detectability, in the interest of future X-ray surveys of the Galactic plane. 

Spectroscopic surveys aimed at detecting infrared counterparts of the Galactic center X-ray population (i.e., red infrared counterparts to hard X-ray sources) will be challenged by the large number of spurious matches. Indeed, of the 2575 infrared matches detected with $K_s\le15.6$ mag, $2347\pm100$ (2$\sigma$) of them are expected to be accidental, and the average probability among these matches is 7\%. In benefit to the pursuit of counterparts, the data in Table 3 provide the means to prioritize spectroscopic targets based on their relative probabilities. In general, the highest probabilities are associated with the brightest infrared matches to the X-ray sources with the smallest positional uncertainties.  Most of the sources with $\sigma_{\textrm{\scriptsize{X}}}\lesssim0\farcs5$ are in the central field near {\sgrastar}, owing to the large number of observations and high photon statistics in that particular region. Repeated deep \textit{Chandra} observations of the extended survey area could feasibly localize the majority of the current catalog of X-ray sources to the level of accuracy achieved in the central field around {\sgrastar}, which would eliminate a large number of spurious infrared matches. Such observations would complement data from the \textit{United Kingdom Infrared Digital Sky Survey} (UKIDSS; Lawrence et al. 2007),  which will produce a deeper catalog, down to $K_s\lesssim17$--18 mag. UKIDSS will certainly produce a large increase in the number of candidate counterparts to X-ray sources, but will also produce a large increase in the number of spurious matches that lie within the error circles of the current \textit{Chandra} catalog. Thus, additional \textit{Chandra} observations are warranted, and would greatly aid in the identification and characterization of the Galactic center X-ray population.

\begin{acknowledgments}
This research was based upon observations made with the \textit{Chandra X-ray Observatory}. Support for this research was provided through \textit{Chandra} award No. GO6-7135I. We thank the referee for a very insightful and helpful critique of this manuscript.
\end{acknowledgments}

\begin{deluxetable}{rrrrrr}
\tablecolumns{6}
\tablewidth{0pc}
\tabletypesize{\scriptsize}
\tablecaption{Statistics of Infrared Matches to Hard X-ray Sources}
\tablehead{\multicolumn{6}{c}{Red Matches} \\   
\colhead{$K_s$} & \colhead{$N_{\textrm{\tiny{obs}}}$} & \colhead{$N_{\textrm{\tiny{ran}}}$} & \colhead{$\sigma$} & \colhead{$N_{\textrm{\tiny{real}}}$}  & \colhead{$N_{\textrm{\tiny{real}}}/\sigma$} 
}
\startdata
 6.0 &        0 &      0 &      \nodata &      \nodata &    \nodata \\
 6.5 &        1 &      0.1 &      0.3 &      0.9 &      2.8 \\
 7.0 &        1 &      0.2 &      0.4 &      0.8 &      1.8 \\
 7.5 &        4 &      0.4 &      0.6 &      3.6 &      5.8 \\
 8.0 &        9 &      0.9 &      0.9 &      8.1 &      8.6 \\
 8.5 &        9 &      1.4 &      1.2 &      7.6 &      6.1 \\
 9.0 &       10 &      3.1 &      1.7 &      6.9 &      4.2 \\
 9.5 &       14 &      9.3 &      3.3 &      4.7 &      1.4 \\
10.0 &       41 &     23.8 &      4.3 &     17.2 &      4.0 \\
10.5 &       76 &     49.0 &      8.1 &     27.0 &      3.3 \\
11.0 &      122 &     85.2 &      9.5 &     36.8 &      3.9 \\
11.5 &      196 &    134.0 &     13.0 &     62.0 &      4.8 \\
12.0 &      261 &    200.2 &     14.6 &     60.8 &      4.2 \\
12.5 &      383 &    293.9 &     18.1 &     89.1 &      4.9 \\
13.0 &      520 &    426.4 &     21.9 &     93.6 &      4.3 \\
13.5 &      730 &    622.9 &     24.8 &    107.1 &      4.3 \\
14.0 &     1012 &    891.9 &     29.5 &    120.1 &      4.1 \\
14.5 &     1497 &   1329.9 &     35.9 &    167.1 &      4.7 \\
15.0 &     2094 &   1893.2 &     39.5 &    200.8 &      5.1 \\
15.6 &     2575 &   2346.6 &     49.7 &    228.4 &      4.6 \\ [2pt]
\hline
\hline \\ [-4pt]
\multicolumn{6}{c}{Blue Matches} \\    
$K_s$ & $N_{\textrm{\tiny{obs}}}$ & $N_{\textrm{\tiny{ran}}}$ & \colhead{$\sigma$} & \colhead{$N_{\textrm{\tiny{real}}}$} & \colhead{$N_{\textrm{\tiny{real}}}/\sigma$} \\ [2pt] 
\hline \\ [-4pt]
 9.0 &        0 &      0.5 &      0.6 &     \nodata &     \nodata \\
 9.5 &        2 &      0.8 &      0.8 &      1.2 &      1.4 \\
10.0 &        2 &      1.8 &      1.3 &      0.2 &      0.2 \\
10.5 &        6 &      3.6 &      1.9 &      2.4 &      1.2 \\
11.0 &       11 &      5.2 &      2.4 &      5.8 &      2.4 \\
11.5 &       17 &      7.6 &      2.8 &      9.4 &      3.4 \\
12.0 &       31 &     11.1 &      3.2 &     19.9 &      6.3 \\
12.5 &       56 &     16.2 &      4.4 &     39.8 &      9.0 \\
13.0 &       74 &     24.0 &      5.2 &     50.0 &      9.6 \\
13.5 &      109 &     35.0 &      5.2 &     74.0 &     14.2 \\
14.0 &      149 &     53.6 &      8.3 &     95.4 &     11.5 \\
14.5 &      198 &     84.3 &      8.2 &    113.7 &     13.9 \\
15.0 &      264 &    120.3 &     11.5 &    143.7 &     12.5 \\
15.6 &      340 &    174.2 &     13.9 &    165.8 &     11.9 \\

\enddata
\tablecomments{$N_{\textrm{\tiny{obs}}}$ is the number of
infrared matches to X-ray sources for aligned catalogs. $N_{\textrm{\tiny{ran}}}$ and $\sigma$ are the average number and standard deviation of matches from 100 random positional offsets.  $N_{\textrm{\tiny{real}}}$ is the estimated number of true counterparts within the given $K_s$ magnitude limit.}
\label{tbl:cpt_stat_hard}
\end{deluxetable}

\begin{deluxetable}{rrrrrr}
\tablecolumns{6}
\tablewidth{0pc}
\tabletypesize{\scriptsize}
\tablecaption{Statistics of Infrared Matches to Soft X-ray Sources}
\tablehead{\multicolumn{6}{c}{Blue Matches} \\   
\colhead{$K_s$} & \colhead{$N_{\textrm{\tiny{obs}}}$} & \colhead{$N_{\textrm{\tiny{ran}}}$} & \colhead{$\sigma$} & \colhead{$N_{\textrm{\tiny{real}}}$}  & \colhead{$N_{\textrm{\tiny{real}}}/\sigma$} 
}
\startdata
 7.0&        0 &      0 &      \nodata &      0 &      \nodata \\
 7.5 &        1 &      0.0 &      0.0 &      1.0 &      0.0 \\
 8.0 &        1 &      0.0 &      0.1 &      1.0 &      9.9 \\
 8.5 &        5 &      0.1 &      0.3 &      4.9 &     18.0 \\
 9.0 &       14 &      0.1 &      0.4 &     13.9 &     37.8 \\
 9.5 &       34 &      0.6 &      0.8 &     33.3 &     44.0 \\
10.0 &       49 &      1.1 &      1.0 &     47.9 &     46.3 \\
10.5 &       63 &      2.1 &      1.2 &     60.9 &     49.4 \\
11.0 &       82 &      3.3 &      1.9 &     78.7 &     41.0 \\
11.5 &      124 &      4.6 &      1.9 &    119.4 &     61.9 \\
12.0 &      163 &      7.1 &      2.7 &    155.9 &     58.1 \\
12.5 &      221 &     11.0 &      2.7 &    210.0 &     76.5 \\
13.0 &      327 &     15.6 &      4.3 &    311.4 &     72.1 \\
13.5 &      440 &     23.4 &      4.2 &    416.6 &     99.2 \\
14.0 &      568 &     36.6 &      6.7 &    531.4 &     79.7 \\
14.5 &      728 &     55.4 &      8.4 &    672.6 &     80.4 \\
15.0 &      879 &     83.6 &      8.9 &    795.4 &     89.3 \\
15.6 &     1007 &    116.9 &     11.6 &    890.0 &     77.0 \\
[2pt]
\hline
\hline \\ [-4pt]
\multicolumn{6}{c}{Red Matches} \\     
$K_s$ & $N_{\textrm{\tiny{obs}}}$ & $N_{\textrm{\tiny{ran}}}$ & \colhead{$\sigma$} & \colhead{$N_{\textrm{\tiny{real}}}$} & \colhead{$N_{\textrm{\tiny{real}}}/\sigma$} \\ [2pt] 
\hline \\ [-4pt]
 8.0 &        0 &      0.3 &      0.5 &     \nodata &     \nodata \\
 8.5 &        1 &      0.7 &      1.0 &      0.3 &      0.4 \\
 9.0 &        3 &      1.6 &      1.3 &      1.4 &      1.1 \\
 9.5 &        8 &      5.0 &      2.1 &      3.0 &      1.4 \\
10.0 &       21 &     14.2 &      4.2 &      6.8 &      1.6 \\
10.5 &       38 &     26.1 &      4.2 &     11.9 &      2.8 \\
11.0 &       56 &     46.0 &      7.1 &     10.0 &      1.4 \\
11.5 &       85 &     69.6 &      8.4 &     15.4 &      1.8 \\
12.0 &      121 &    105.0 &     10.8 &     16.0 &      1.5 \\
12.5 &      180 &    154.9 &     12.2 &     25.1 &      2.1 \\
13.0 &      247 &    222.0 &     17.1 &     25.0 &      1.5 \\
13.5 &      333 &    326.5 &     16.8 &      6.5 &      0.4 \\
14.0 &      492 &    465.1 &     24.4 &     26.9 &      1.1 \\
14.5 &      700 &    717.8 &     27.4 &    -17.8 &     $-$0.6 \\
15.0 &      998 &   1058.2 &     36.6 &    -60.2 &     $-$1.6 \\
15.6 &     1262 &   1373.6 &     37.2 &   -111.6 &    $-$3.0 \\
 \enddata
\tablecomments{Columns are the same as in Table 1.}
\label{tbl:cpt_stat_soft}
\end{deluxetable}

\begin{landscape}
\setlength{\tabcolsep}{0.05in}
\renewcommand{\arraystretch}{0.90}
\tiny
\begin{center}
\begin{deluxetable}{lcccccccccrrrr}
\tablecolumns{14}
\tablewidth{0pc}
\tabletypesize{\scriptsize}
\tablecaption{\normalsize{Candidate Infrared Counterparts to X-ray Sources Toward the Galactic Center}}
\tablehead{

\colhead{ID} & \colhead{ID$_{\textrm{\tiny{X}}}$} & \colhead{X-ray Source} &  \colhead{$\sigma_{\textrm{\tiny{X}}}$}  & \colhead{Type$_{\textrm{\tiny{X}}}$} & \colhead{Type$_{\textrm{\tiny{IR}}}$} & \colhead{R.A.$_{\textrm{\tiny{IR}}}$} & \colhead{Dec.$_{\textrm{\tiny{IR}}}$} & \colhead{$\phi$} & \colhead{$\rho(K_s)$} &\colhead{$P$} & \colhead{$J$}  & \colhead{$H$}  & \colhead{$K_s$}   \\ [2pt]
\colhead{} & \colhead{} &  \colhead{(CXOUGC J)} &  \colhead{(arcsec)} & \multicolumn{2}{c}{} & \multicolumn{2}{c}{(degrees, J2000)} & \colhead{(arcsec)} & \colhead{(arcsec$^{-2}$)}& \colhead{(\%)}  & \colhead{(mag)} & \colhead{(mag)} & \colhead{(mag)}   \\ [4pt]
\colhead{(1)} & \colhead{(2)} & \colhead{(3)} & \colhead{(4)} & \colhead{(5)} & \colhead{(6)} & \colhead{(7)} & \colhead{(8)} & \colhead{(9)} & \colhead{(10)} & \colhead{(11)} & \colhead{(12)} & \colhead{(13)} &\colhead{(14)} 
}
\startdata
   1&    1& 174457.1$-$285740& 1.8& soft& blue& 266.23814& $-$28.96102& 0.69& 0.0249&   42.2&$   14.60\pm0.02$&$   13.91\pm0.02$&$   13.89\pm0.06$\\
   2&    1& 174457.1$-$285740& 1.8& soft&  red& 266.23841& $-$28.96159& 1.65& 0.0519&\nodata&        \nodata  &$   16.68\pm0.11$&$   14.99\pm0.09$\\
   3&    4& 174459.9$-$290324& 2.0& hard& blue& 266.24973& $-$29.05640& 1.57& 0.0010&   15.4&$   14.90\pm0.02$&$   13.69\pm0.01$&$   12.90\pm0.03$\\
   4&    4& 174459.9$-$290324& 2.0& hard& blue& 266.25009& $-$29.05688& 0.88& 0.0054&   12.0&        \nodata  &$   16.19\pm0.07$&$   15.41\pm0.10$\\
   5&    4& 174459.9$-$290324& 2.0& hard&  red& 266.24967& $-$29.05680& 0.47& 0.0790&    1.2&        \nodata  &$   16.08\pm0.07$&$   14.60\pm0.06$\\
   6&    4& 174459.9$-$290324& 2.0& hard&  red& 266.24974& $-$29.05726& 1.57& 0.0441&    1.2&        \nodata  &$   15.26\pm0.03$&$   13.90\pm0.03$\\
   7&    5& 174459.9$-$290538& 1.5& hard&  red& 266.25005& $-$29.09386& 1.10& 0.0795&    2.0&        \nodata  &$   16.83\pm0.15$&$   14.86\pm0.11$\\
   8&    6& 174500.2$-$290057& 2.6& hard&  red& 266.25120& $-$29.01551& 1.69& 0.0060&    1.8&$   16.00\pm0.04$&$   12.83\pm0.02$&$   11.23\pm0.01$\\
   9&    6& 174500.2$-$290057& 2.6& hard&  red& 266.25079& $-$29.01604& 1.08& 0.0209&    1.5&        \nodata  &$   14.44\pm0.02$&$   12.82\pm0.04$\\
  10&    6& 174500.2$-$290057& 2.6& hard&  red& 266.25130& $-$29.01645& 1.80& 0.0223&    1.2&$   17.27\pm0.08$&$   14.22\pm0.02$&$   12.93\pm0.02$\\
  11&    6& 174500.2$-$290057& 2.6& hard&  red& 266.25071& $-$29.01643& 2.09& 0.0629&    0.5&        \nodata  &$   15.62\pm0.07$&$   14.16\pm0.09$\\
  12&   10& 174501.7$-$290313& 1.1& hard&  red& 266.25738& $-$29.05380& 0.48& 0.0208&    7.7&        \nodata  &$   15.23\pm0.04$&$   13.05\pm0.02$\\
  13&   13& 174502.2$-$285749& 1.0& hard&  red& 266.25935& $-$28.96376& 0.41& 0.0019&   29.8&$   14.08\pm0.02$&$   11.93\pm0.01$&$   10.81\pm0.01$\\
  14&   14& 174502.4$-$290205& 1.1& soft& blue& 266.26005& $-$29.03490& 0.08& 0.0275&   58.4&$   13.83\pm0.01$&$   13.31\pm0.02$&$   13.26\pm0.03$\\
  15&   15& 174502.4$-$290453& 1.4& soft&  red& 266.26018& $-$29.08172& 0.81& 0.0665&\nodata&        \nodata  &        \nodata  &$   14.68\pm0.11$\\
  16&   18& 174502.8$-$290429& 1.3& soft& blue& 266.26195& $-$29.07482& 0.10& 0.0151&   53.5&$   13.36\pm0.02$&$   12.88\pm0.01$&$   12.75\pm0.04$\\
  17&   19& 174502.9$-$285920& 1.1& hard&  red& 266.26227& $-$28.98982& 0.54& 0.0348&    5.4&        \nodata  &$   15.21\pm0.14$&$   13.48\pm0.07$\\
  18&   21& 174503.8$-$290004& 0.9& hard&  red& 266.26600& $-$29.00125& 0.71& 0.0783&    4.4&        \nodata  &$   15.71\pm0.08$&$   14.32\pm0.09$\\
  19&   22& 174504.1$-$285902& 4.7& hard& blue& 266.26711& $-$28.98410& 0.95& 0.0035&    4.1&        \nodata  &$   16.07\pm0.07$&$   15.21\pm0.09$\\
  20&   22& 174504.1$-$285902& 4.7& hard& blue& 266.26628& $-$28.98387& 3.53& 0.0034&    3.3&        \nodata  &$   15.97\pm0.10$&$   15.12\pm0.07$\\
  21&   22& 174504.1$-$285902& 4.7& hard&  red& 266.26713& $-$28.98368& 1.44& 0.0201&    0.5&        \nodata  &$   14.43\pm0.02$&$   12.88\pm0.02$\\
  22&   22& 174504.1$-$285902& 4.7& hard&  red& 266.26694& $-$28.98485& 3.37& 0.0611&    0.2&        \nodata  &$   16.01\pm0.06$&$   14.34\pm0.05$\\
  23&   22& 174504.1$-$285902& 4.7& hard&  red& 266.26710& $-$28.98330& 2.70& 0.0843&    0.2&        \nodata  &$   16.21\pm0.07$&$   14.80\pm0.06$\\
  24&   22& 174504.1$-$285902& 4.7& hard&  red& 266.26788& $-$28.98384& 1.65& 0.0970&    0.2&        \nodata  &$   16.10\pm0.04$&$   15.05\pm0.07$\\
  25&   22& 174504.1$-$285902& 4.7& hard&  red& 266.26831& $-$28.98324& 3.97& 0.0436&    0.2&        \nodata  &$   15.47\pm0.05$&$   13.93\pm0.03$
  \enddata
\tablecomments{~List of candidate infrared counterparts to X-ray sources in the foreground toward the Galactic center. Additional X-ray data on the associated \textit{Chandra} sources can be obtained by using the No.$_{\textrm{\tiny{X}}}$ record numbers to look up the appropriate sources in the tables of Muno et al. (2009), which are included in Column 2. \\ [2pt]
The table columns are as follows: \\
(1) ID number for this work. \\
(2) Record ID in the master catalog of Muno et al. (2009). \\
(3) Source ID for the X-ray sources. Although approximately correct, these are not necessarily equivalent
to the actual X-ray positions, which are given in Muno et al. (2009). \\
(4) Uncertainty of the position of the X-ray source (error circle) in arcseconds ($\sigma_{\textrm{\tiny{X}}}$).\\
(5) X-ray source hardness designation, based on HR0 values (see Section 2.2). \\
(6) Infrared color designation (see Section 2.1). \\
(7-8) J2000 R.A. and decl. of the infrared sources. \\
(9) Separation between the X-ray and infrared source in arcseconds ($\phi$). \\
(10) Percent probability that the infrared match is the genuine counterpart, calculated using equations 2 and 3. \\
(11) Local density of infrared sources, $\rho (K_s)$, having $K_s$ values equivalent to, or greater than, the infrared match in question. \\
(12--14) Magnitudes of the infrared matches in the \textit{J}, \textit{H} and $K_s$ bands.} \\
This table is available in its entirety in a machine-readable form in the online journal. A portion is shown here for guidance regarding its form and content. 
\label{tab:nir_ctps}
\end{deluxetable}
\end{center}
\end{landscape}

\begin{deluxetable}{rcrrcc}
\tablecolumns{6}
\tablewidth{0pc}
\tabletypesize{\scriptsize}
\tablecaption{Known Massive Stellar X-ray Sources}
\tablehead{
\colhead{X-ray Source} & \colhead{$\sigma_{\textrm{\tiny{X}}}$} & \colhead{$K_s$} & \colhead{$P$} & \colhead{Association} &  \colhead{Ref.} \\
\colhead{(CXOGC J)} & \colhead{(arcsec)}  & \colhead{(mag)} & \colhead{(\%)} & \colhead{} & \colhead{} }
\startdata
174516.1$-$290315 & 0.4 & $7.89$ & 79.8 & Field &  7 \\
174528.6$-$285605 & 0.5 & $9.72$ & 41.4 & Field   & 1 \\
174536.1$-$285638 & 0.4 & $10.42$ & 70.6 & Field & 6 \\
174549.7$-$284925 & 0.4 & $12.16$  & 31.8 & Arches Member 2  & 3 \\
174550.2$-$284911 & 0.3 & $10.47$ & 70.9 & Arches Member 9 & 3 \\
174550.4$-$284922 & 0.3 & $9.87$ & 84.0 & Arches Member 6 & 3 \\
174550.4$-$284919 & 0.3 & $9.56$ & 73.5 & Arches Member 7 &  3 \\
174550.6$-$285919 & 0.4 & $10.85$ & \nodata & Field & 1 \\
174555.3$-$285126 & 0.5 & $10.97$& 41.0& Field & 5 \\
174614.6$-$284940 & 0.6 & $7.29$ & 26.8 & Quintuplet Member 231 & 2  \\
174615.8$-$284945 & 0.5 & $7.24$ & 25.9 & Quintuplet Member 211 & 2  \\
174616.6$-$284909 & 0.8 & $11.08$ &  13.7 &   Quintuplet Member 344 & 2  \\
174617.0$-$285131 & 0.6 & $10.49$ & 31.9 & Field & 5 \\
174645.2$-$281547 & 1.0 & $7.18$ & 16.2 & Field & 4
 \enddata
 \tablecomments{The table includes known, X-ray-emitting Wolf-Rayet and O-supergiant stars  recovered by our cross-corrrelation, which are also included in Table 3. Additional data on these stars will be presented in a forthcoming paper. References for stellar identifications: (1) Cotera et al. (1999); (2) Figer et al. (1999); (3) Figer et al. (2002), and references therein; (4) Hyodo et al. (2008); (5) Mauerhan et al. (2007); (6) Mikles et al. (2006); (7) Muno et al. (2006).}
\label{tbl:cpt_stat_soft}
\end{deluxetable}

\begin{deluxetable}{rrrrrr}
\tablecolumns{6}
\tablewidth{0pc}
\tabletypesize{\scriptsize}
\tablecaption{Statistics of Infrared Matches in Subregions (1)}
\tablehead{\multicolumn{6}{c}{{Near Sgr A}} \\   
\colhead{$K_s$} & \colhead{$N_{\textrm{\tiny{obs}}}$} & \colhead{$N_{\textrm{\tiny{ran}}}$} & \colhead{$\sigma$} & \colhead{$N_{\textrm{\tiny{real}}}$}  & \colhead{$N_{\textrm{\tiny{real}}}/\sigma$} 
}
\startdata
 7.5 &      0 &      0.1 &      0.3 &     0 &     \nodata \\
 8.0 &      4 &      0.3 &      0.7 &      3.7 &      5.1 \\
 8.5 &      4 &      0.3 &      0.6 &      3.7 &      6.1 \\
 9.0 &      4 &      0.9 &      1.1 &      3.1 &      2.8 \\
 9.5 &      6 &      2.2 &      1.5 &      3.8 &      2.5 \\
10.0 &     16 &      7.1 &      2.3 &      8.9 &      3.9 \\
10.5 &     27 &     15.0 &      4.0 &     12.0 &      3.0 \\
11.0 &     38 &     28.3 &      4.8 &      9.7 &      2.0 \\
11.5 &     57 &     44.7 &      6.3 &     12.3 &      2.0 \\
12.0 &     79 &     66.2 &      8.4 &     12.8 &      1.5 \\
12.5 &    107 &     97.7 &     10.2 &      9.3 &      0.9 \\
13.0 &    153 &    139.7 &     11.8 &     13.3 &      1.1 \\
13.5 &    220 &    200.8 &     14.3 &     19.2 &      1.3 \\
14.0 &    307 &    285.6 &     17.9 &     21.4 &      1.2 \\
14.5 &    429 &    393.4 &     18.8 &     35.6 &      1.9 \\
15.0 &    545 &    507.0 &     21.3 &     38.0 &      1.8 \\
15.6 &    617 &    570.8 &     23.2 &     46.2 &      2.0 \\
[2pt]
\hline
\hline \\ [-4pt]
\multicolumn{6}{c}{Near the Arches and Quintuplet} \\     
$K_s$ & $N_{\textrm{\tiny{obs}}}$ & $N_{\textrm{\tiny{ran}}}$ & \colhead{$\sigma$} & \colhead{$N_{\textrm{\tiny{real}}}$} & \colhead{$N_{\textrm{\tiny{real}}}/\sigma$} \\ [2pt] 
\hline \\ [-4pt]
7.0 &      0 &      0 &      \nodata &      0 &      \nodata \\
 7.5 &      2 &      0.1 &      0.2 &      1.9 &      8.1 \\
 8.0 &      3 &      0.1 &      0.4 &      2.9 &      7.8 \\
 8.5 &      3 &      0.2 &      0.4 &      2.8 &      7.7 \\
 9.0 &      3 &      0.4 &      0.6 &      2.6 &      4.4 \\
 9.5 &      3 &      0.9 &      0.9 &      2.1 &      2.3 \\
10.0 &      8 &      2.2 &      1.9 &      5.8 &      3.0 \\
10.5 &     10 &      5.1 &      2.3 &      4.9 &      2.2 \\
11.0 &     14 &      8.2 &      3.1 &      5.8 &      1.9 \\
11.5 &     21 &     12.7 &      3.8 &      8.3 &      2.2 \\
12.0 &     26 &     19.2 &      4.7 &      6.8 &      1.4 \\
12.5 &     42 &     28.1 &      5.4 &     13.9 &      2.6 \\
13.0 &     61 &     42.4 &      7.5 &     18.6 &      2.5 \\
13.5 &     80 &     61.5 &      7.9 &     18.5 &      2.3 \\
14.0 &    107 &     89.6 &     10.0 &     17.4 &      1.7 \\
14.5 &    151 &    131.9 &     11.5 &     19.1 &      1.7 \\
15.0 &    199 &    181.8 &     14.3 &     17.2 &      1.2 \\
15.6 &    239 &    215.1 &     13.9 &     23.9 &      1.7 \\
 \enddata
\tablecomments{Columns are the same as in Table 1.}
\label{tbl:cpt_stat_soft}
\end{deluxetable}

\begin{deluxetable}{rrrrrr}
\tablecolumns{6}
\tablewidth{0pc}
\tabletypesize{\scriptsize}
\tablecaption{Statistics of Infrared Matches in Subregions (2)}
\tablehead{\multicolumn{6}{c}{Sgr B} \\   
\colhead{$K_s$} & \colhead{$N_{\textrm{\tiny{obs}}}$} & \colhead{$N_{\textrm{\tiny{ran}}}$} & \colhead{$\sigma$} & \colhead{$N_{\textrm{\tiny{real}}}$}  & \colhead{$N_{\textrm{\tiny{real}}}/\sigma$} 
}
\startdata
  9.0 &      0 &      0.1 &      0.2 &     0 &     \nodata \\
 9.5 &      1 &      0.3 &      0.6 &      0.7 &      1.2 \\
10.0 &      4 &      0.6 &      0.8 &      3.4 &      4.0 \\
10.5 &      6 &      1.2 &      1.1 &      4.8 &      4.3 \\
11.0 &     11 &      1.8 &      1.6 &      9.2 &      5.8 \\
11.5 &     16 &      2.8 &      1.7 &     13.2 &      7.9 \\
12.0 &     19 &      3.9 &      2.1 &     15.1 &      7.0 \\
12.5 &     23 &      6.0 &      2.4 &     17.0 &      7.1 \\
13.0 &     25 &      8.6 &      3.3 &     16.4 &      5.0 \\
13.5 &     30 &     13.2 &      3.2 &     16.8 &      5.2 \\
14.0 &     39 &     18.4 &      4.2 &     20.6 &      4.9 \\
14.5 &     47 &     29.9 &      4.9 &     17.1 &      3.5 \\
15.0 &     57 &     45.1 &      6.3 &     11.9 &      1.9 \\
15.6 &     74 &     65.4 &      7.9 &      8.6 &      1.1 \\
 [2pt]
\hline
\hline \\ [-4pt]
\multicolumn{6}{c}{Sgr C} \\     
$K_s$ & $N_{\textrm{\tiny{obs}}}$ & $N_{\textrm{\tiny{ran}}}$ & \colhead{$\sigma$} & \colhead{$N_{\textrm{\tiny{real}}}$} & \colhead{$N_{\textrm{\tiny{real}}}/\sigma$} \\ [2pt] 
\hline \\ [-4pt]
 6.5 &      1 &      0 &      0 &      0 &      \nodata\\
 6.5 &      1 &      0.0 &      0.0 &      1.0 &      `$\infty$'\\
 7.0 &      1 &      0.0 &      0.1 &      1.0 &      7.0 \\
 7.5 &      1 &      0.0 &      0.1 &      1.0 &      9.9 \\
 8.0 &      1 &      0.0 &      0.2 &      1.0 &      4.9 \\
 8.5 &      1 &      0.1 &      0.2 &      0.9 &      4.3 \\
 9.0 &      1 &      0.1 &      0.3 &      0.9 &      2.7 \\
 9.5 &      1 &      0.4 &      0.6 &      0.6 &      0.9 \\
10.0 &      1 &      1.1 &      1.1 &     -0.1 &     $-$0.1\\
10.5 &      2 &      2.2 &      1.4 &     -0.2 &     $-$0.2 \\
11.0 &      5 &      4.3 &      2.2 &      0.7 &      0.3 \\
11.5 &      6 &      6.7 &      2.8 &     -0.7 &     $-$0.2 \\
12.0 &      7 &     11.1 &      3.1 &     -4.2 &     $-$1.3 \\
12.5 &     14 &     16.4 &      4.1 &     -2.3 &     $-$0.6 \\
13.0 &     18 &     23.9 &      4.7 &     -5.9 &     $-$1.2 \\
13.5 &     32 &     34.4 &      5.6 &     -2.4 &     $-$0.4 \\
14.0 &     46 &     49.2 &      6.8 &     -3.2 &     $-$0.5 \\
14.5 &     79 &     70.8 &      7.5 &      8.2 &      1.1 \\
15.0 &    118 &    106.1 &      8.1 &     11.9 &      1.5 \\
 \enddata
\tablecomments{Columns are the same as in Table 1.}
\label{tbl:cpt_stat_soft}
\end{deluxetable}

\clearpage
\newpage

\begin{landscape}
\begin{figure*}[h]
\centering
\epsscale{1.0}
\plottwo{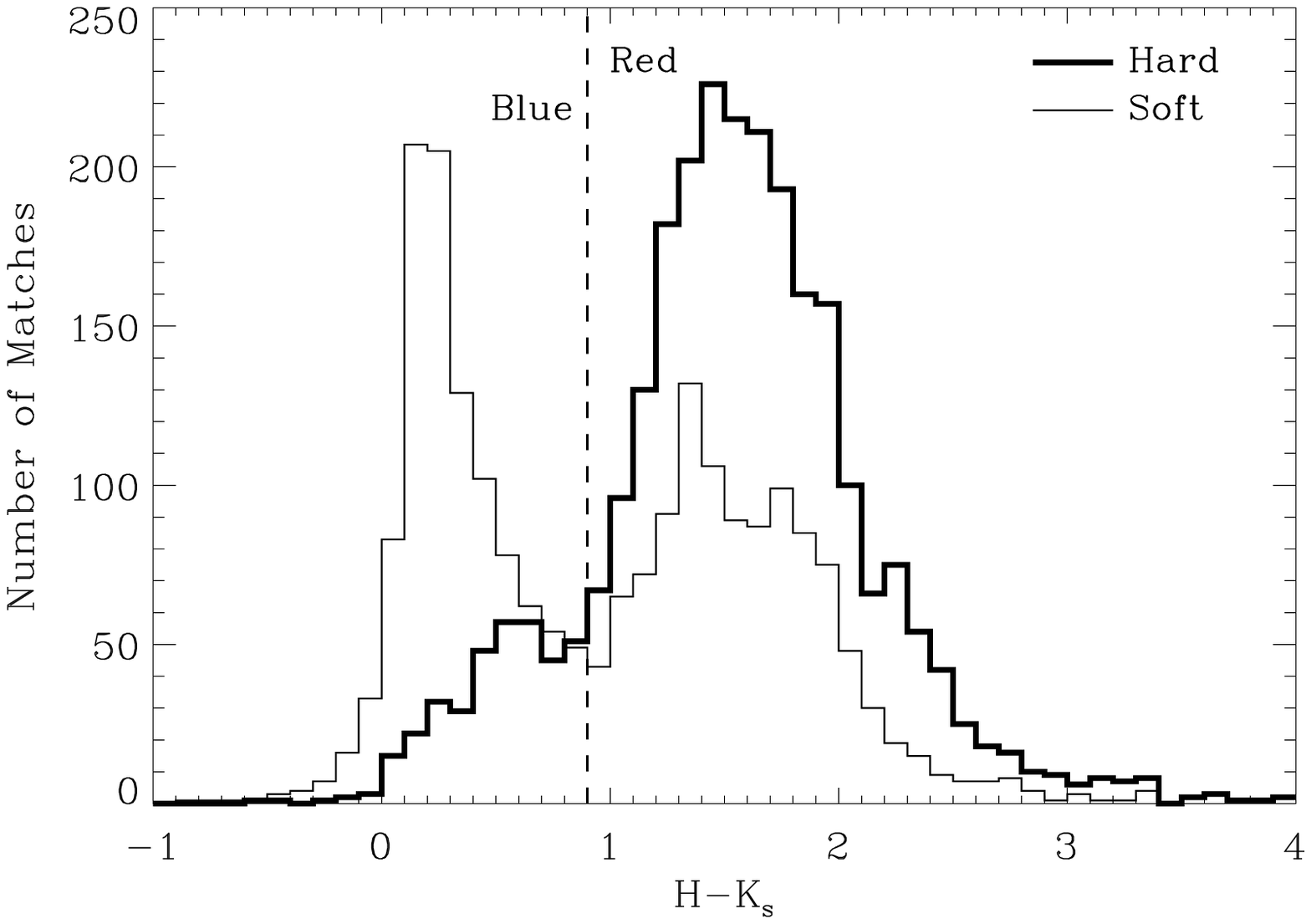}{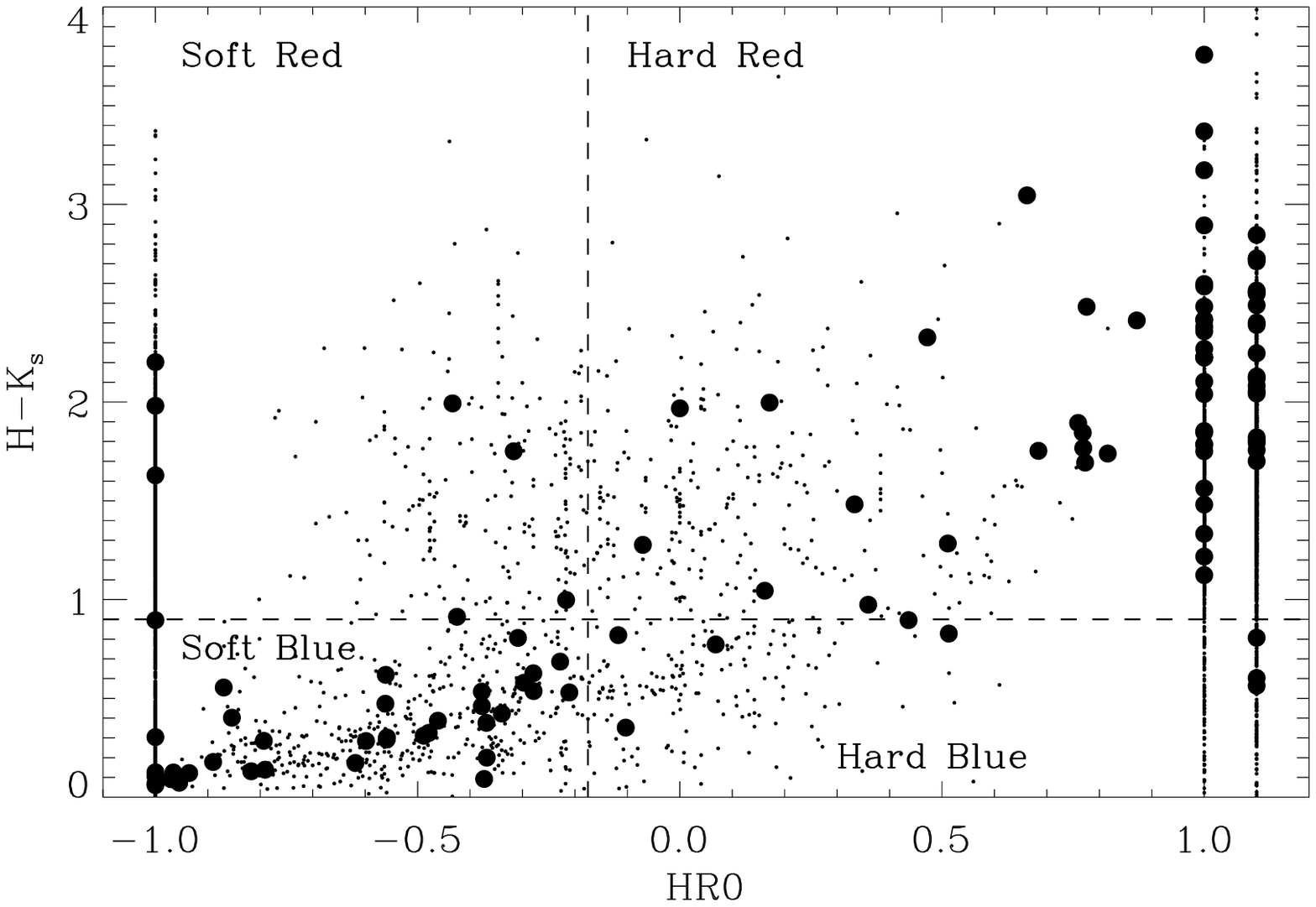}
\caption[]{\linespread{1}\normalsize{\textit{Left panel: }$H-K_s$ color histograms for all infrared matches to hard (\textit{thick line}) and soft (\textit{thin line}) X-ray sources toward the Galactic center. \textit{Right panel: } scatter plot of soft X-ray color, HR0, and $H-K_s$ for infrared matches to X-ray sources (\textit{small dots}). Matches not detected in any of the soft X-ray bands ($E<3.3$ keV) were assigned $\textrm{HR0}=1.1$ for this figure. Matches with $\sigma_{\textrm{\scriptsize{X}}}\le0\farcs5$ and $K_s<13$ mag are represented by \textit{large dots}; owing to their relative brightness and excellent X-ray astrometry, these matches have the highest probability of being genuine counterparts, and their distribution in the figure justifies our definition of hard and soft X-ray sources, and blue and red infrared sources.}}
\end{figure*}
\end{landscape}

\begin{figure*}[h]
\centering
\epsscale{0.7}
\plotone{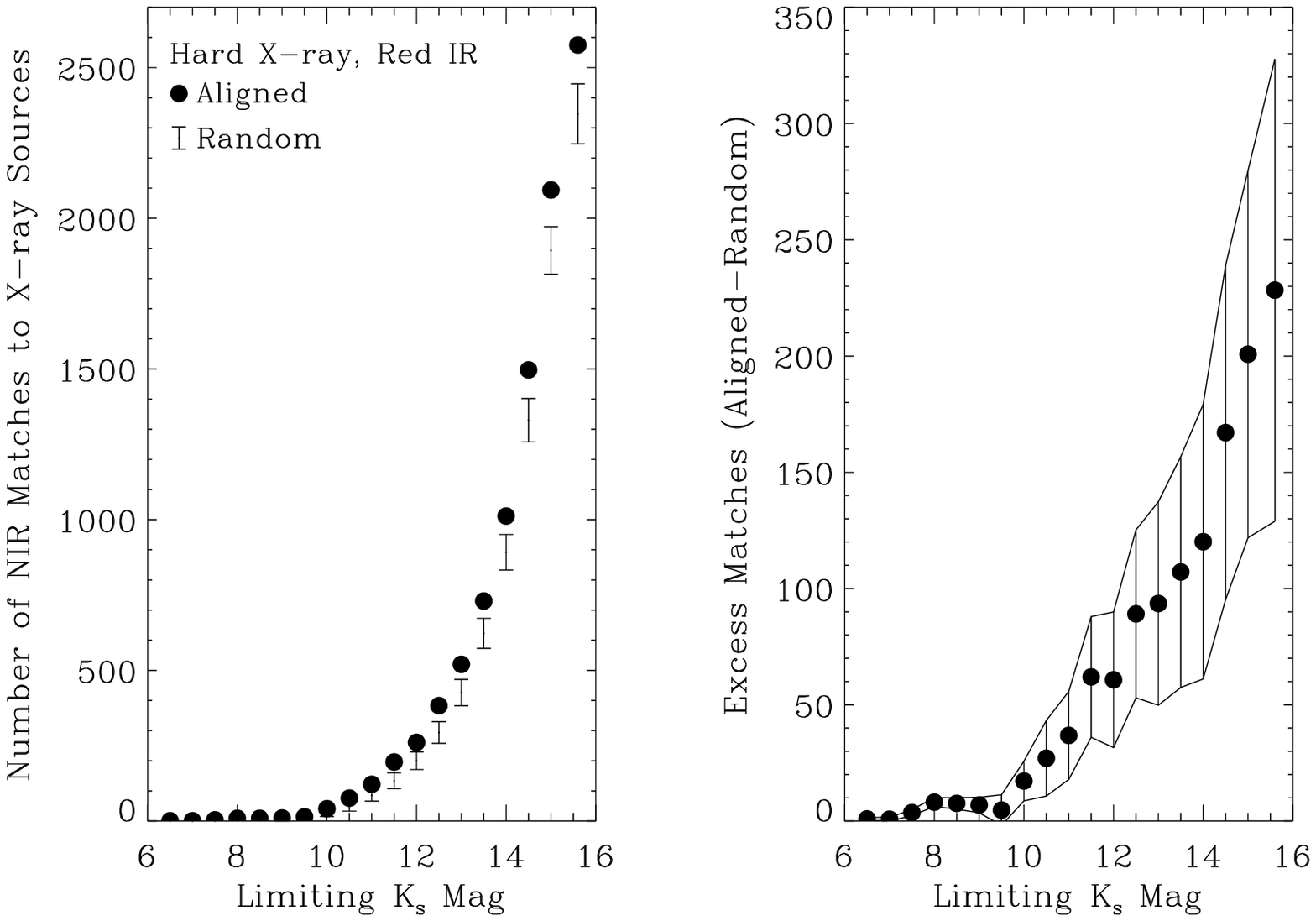}
\plotone{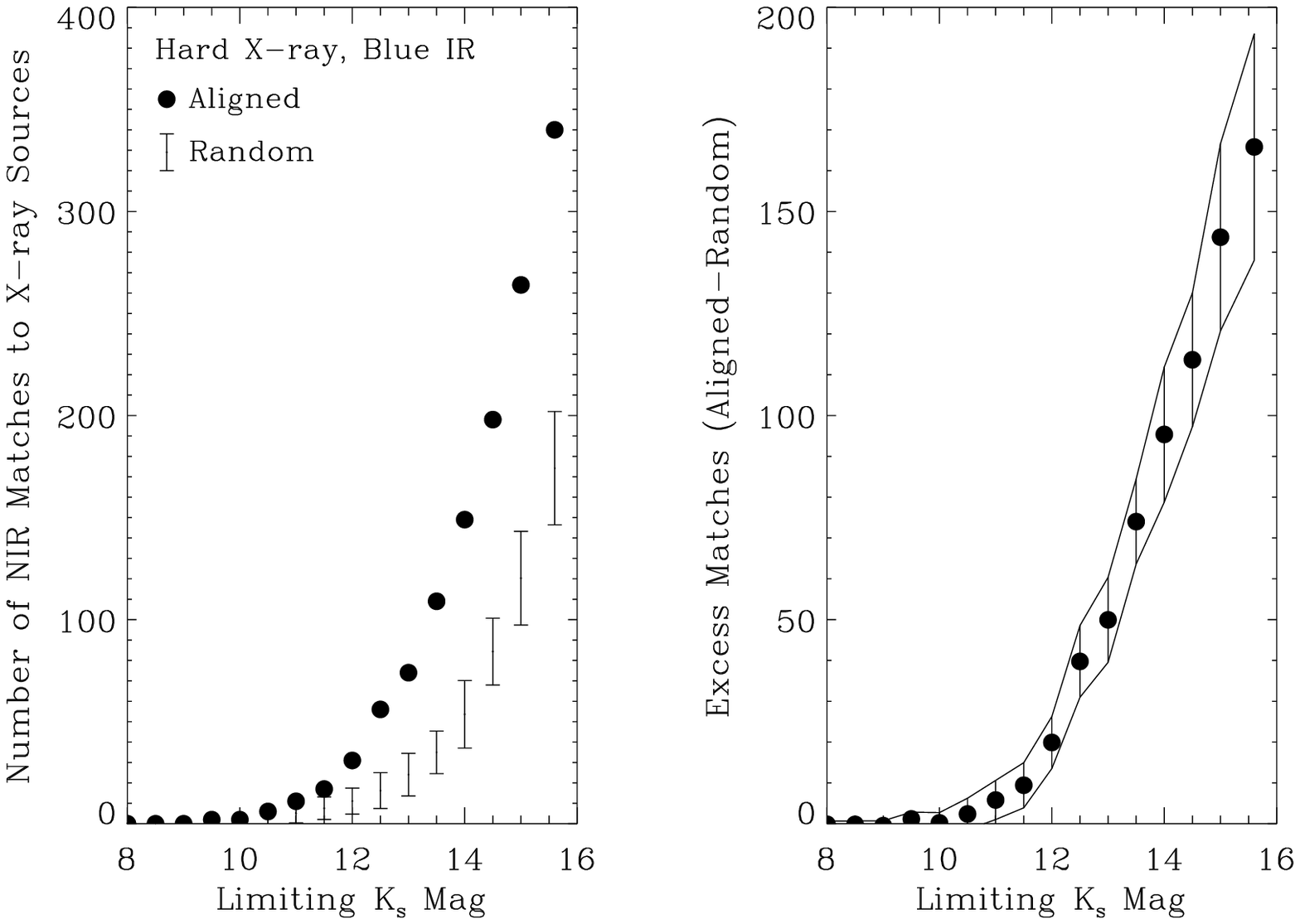}
\caption[]{\linespread{1}\normalsize{Number of infrared matches to hard X-ray sources as a function of limiting $K_s$ magnitude, illustrating the data values listed in Tables 1 and 2. Red sources (upper plots) and blue sources (lower plots) were selected based upon their $H-K_s$ values, which are consistent with distances near the Galactic center and foreground, respectively.  The left panels of the upper and lower plots compare the number of matches detected in aligned catalogs (\textit{filled circles}) with the average number of matches over 100 random offsets in R.A. and decl. (2$\sigma$ error bars), as a function of limiting $K_s$ magnitude. The right panels of the upper and lower plots show the excess number of infrared matches when catalogs are aligned vs. when they are randomly offset (with 2$\sigma$ error bars), which indicates a significant detection of real counterparts.}}
\end{figure*}

\begin{figure*}[h]
\centering
\epsscale{0.7}
\plotone{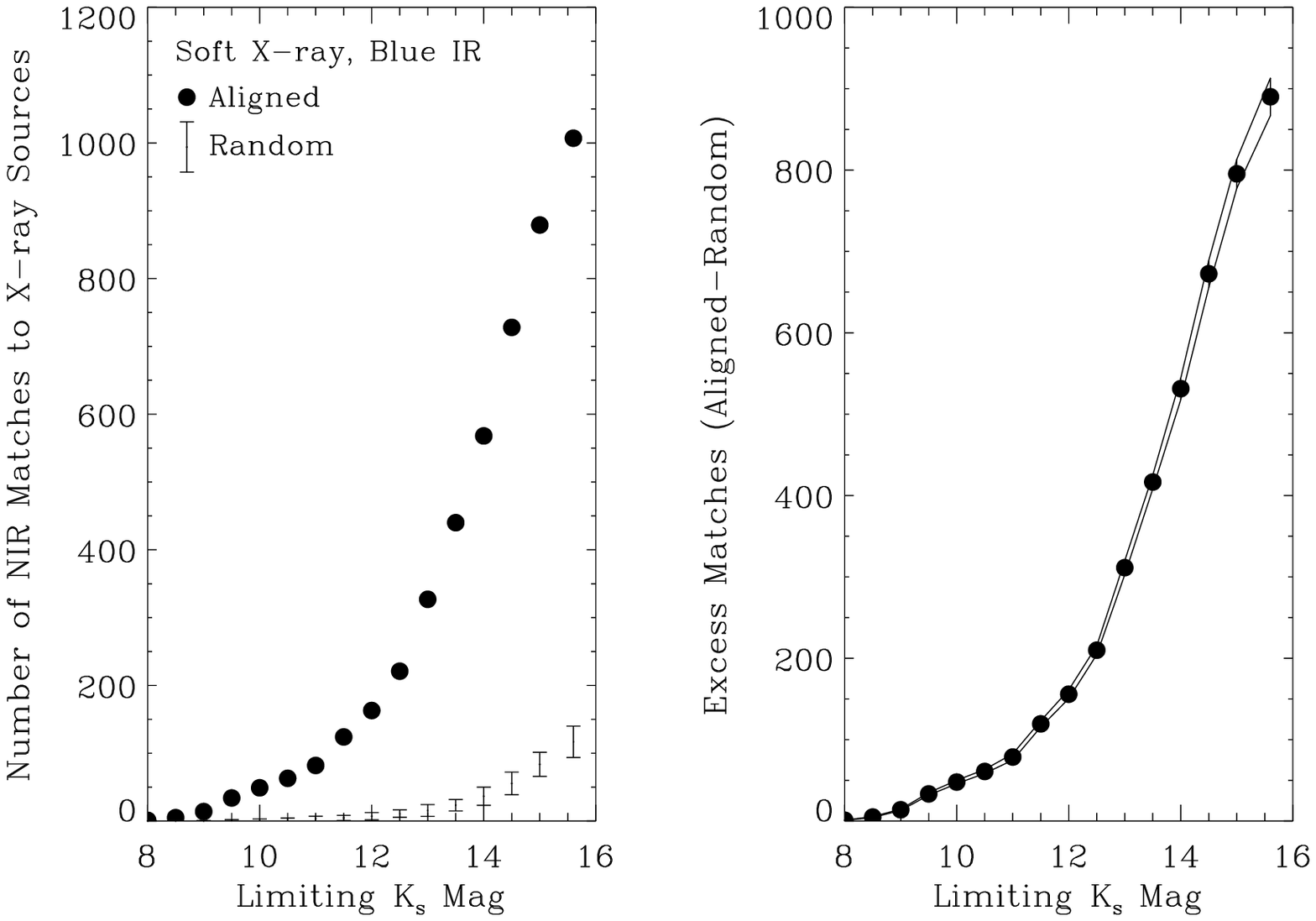}
\plotone{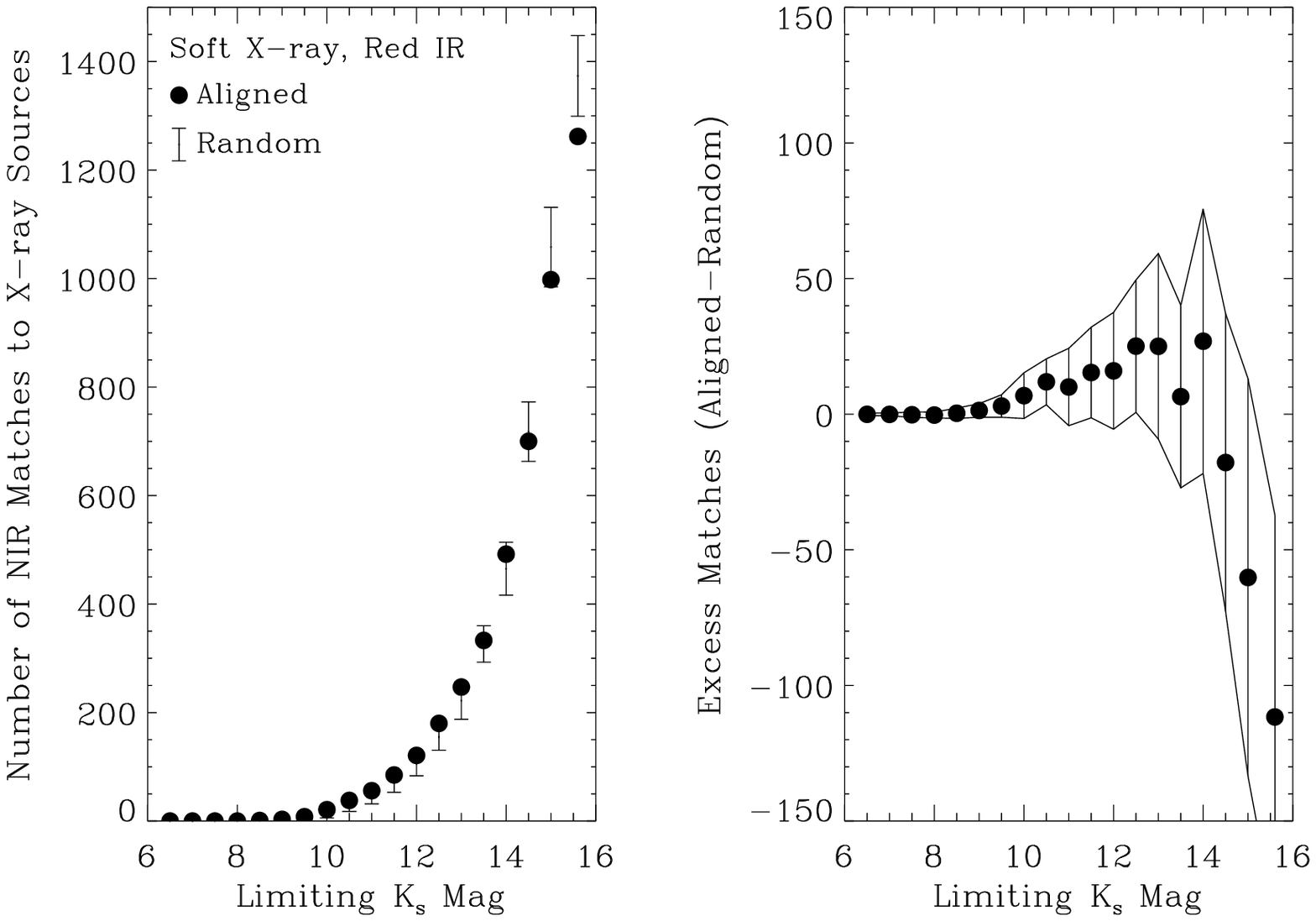}
\caption[]{\linespread{1}\normalsize{Number of infrared matches to soft X-ray sources as a function of limiting $K_s$ magnitude, illustrating the data values listed in Table 2. The illustrated properties are the same as those described in Figure 2.}}
\end{figure*}

\begin{figure*}[h]
\centering
\epsscale{0.9}
\plotone{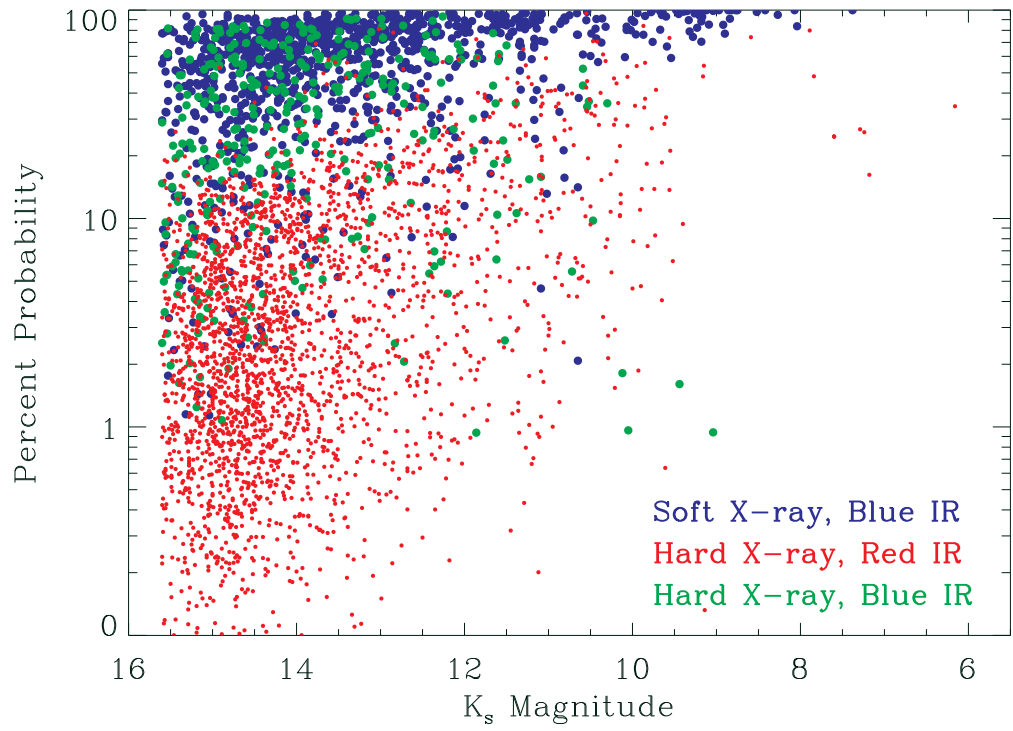}
\caption[]{\linespread{1}\normalsize{Probabilities of infrared matches from Table 3, plotted with respect to $K_s$ magnitude.}}
\end{figure*}

\begin{figure*}[h]
\centering
\epsscale{0.9}
\plotone{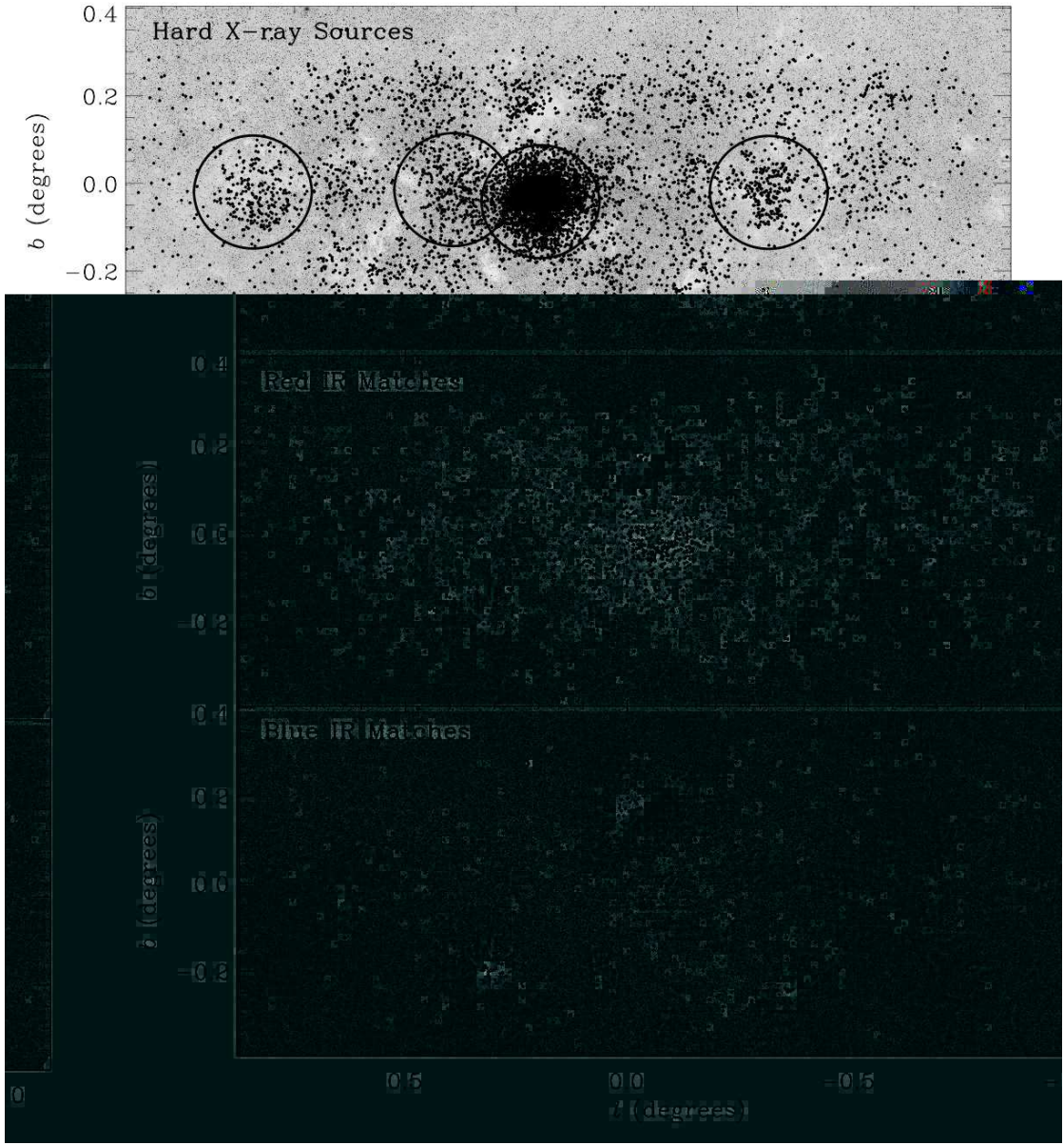}
\caption[]{\linespread{1}\normalsize{Spatial distribution of hard X-ray sources and red and blue infrared matches, plotted on inverted 2MASS $K_s$-band images of the survey area, which covers $l \times b \approx 2\fdg0 \times 0\fdg8$. The circles in the upper panel enclose regions where we performed separate cross-correlation experiments; from left to right: Sgr B, the Arches and Quintuplet region, Sgr A, and Sgr C.}}
\end{figure*}

\begin{figure*}[h]
\centering
\epsscale{0.9}
\plotone{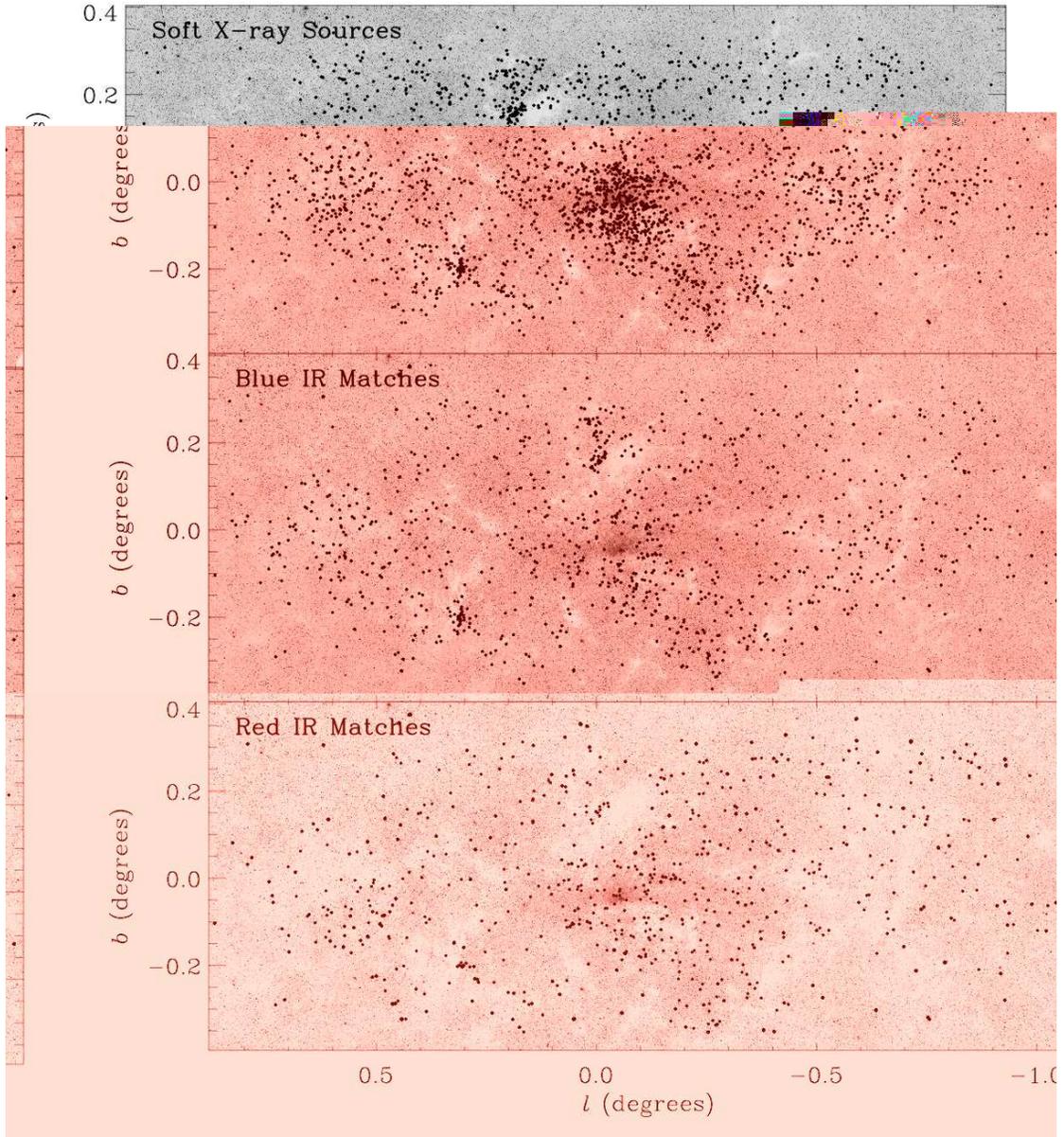}
\caption[]{\linespread{1}\normalsize{Spatial distribution of soft X-ray sources and red and blue infrared matches, plotted on inverted 2MASS $K_s$-band images of the survey area.}}
\end{figure*}

\end{document}